\documentclass[12pt,preprint]{aastex}%AASMS

\def\H#1{H$_2$}
\def\HI{\protect\ion{H}{1}}
\def\HII{\protect\ion{H}{2}}

\def\NI{\protect\ion{N}{1}}

\def\OI{\protect\ion{O}{1}}
\def\OVI{\protect\ion{O}{6}}

\def\AlII{\protect\ion{Al}{2}}
\def\SiII{\protect\ion{Si}{2}}
\def\SII{\protect\ion{S}{2}}
\def\PII{\protect\ion{P}{2}}
\def\ArI{\protect\ion{Ar}{1}}

\def\FeII{\protect\ion{Fe}{2}}
\def\NiII{\protect\ion{Ni}{2}}

\def\FUSE{{\it FUSE}}
\def\HST{{\it HST}}
\def\STIS{{\it STIS}}

\def\dex#1{10$^{#1}$}
\def\tdex#1{$\times$10$^{#1}$}
\def\cmm#1{cm$^{-#1}$}
\def\kms{{\rm km\,s$^{-1}$}}
\def\fu{erg\,\cmm2\,s$^{-1}$\,\AA$^{-1}$}
\def\deg{$^\circ$}
\def\ltsim{<}
\def\gtsim{>}

\def\vlsr{$v_{\rm LSR}$}
\def\vlsra{$\vert$\vlsr$\vert$}
\def\babs{$\vert$$b$$\vert$}
\def\T#1#2{{\it T}$_{#1#2}$}
\def\l{~$\lambda$}
\def\ll{~$\lambda\lambda$}

\def\finalnumberofsources{73}
\def\finaldetections{65}
\def\finalnondetections{8}
\def\ivdetections{16}

\def\ivlos{35}

\def\hvlos{19}
\def\snlimit{10}
\def\Sect#1{Sect.~#1}

\def\Sobserve{2}

\def\Sshift{2.2}
\def\Smeasure{3}
\def\Sresults{4}

\def\Sspecial{4.2}

\def\SIVC{4.4}

\def\Sexternal{4.6}
\def\Scorrel{4.7}

\def\Tobservations{1}
\def\Tvelocities{2}
\def\Tmolhyd{3}
\def\TIVC{4}
\def\Fexamples{1}
\def\Fspecial{2}
\def\Ftset{3}
\def\FHIHt{4}
\def\Fexternal{5}
\def\Fcorrel{6}
\def\FskymapN{7}
\def\FskymapS{8}

\begin{document}

\title{A FUSE survey of high-latitude Galactic molecular hydrogen}
\author{B.P. Wakker\altaffilmark{1}}
\altaffiltext{1}{Department of Astronomy, University of Wisconsin, 475 N. Charter St, Madison, WI 53706.
Based on observations made with the NASA-CNES-CSA Far Ultraviolet Spectroscopic
Explorer. \FUSE\ is operated for NASA by the Johns Hopkins University under NASA
contract NAS5-32985.}

%%%%%%%%%%%%%%%%%%%%%%%%%%%%%%%%%%%%%%%%%%%%%%%%%%%%%%%%%%%%%%%%%%%%%%%%%%%%%%%%

\begin{abstract}
Measurements of molecular hydrogen (\H2) column densities are presented for
the first six rotational levels ($J$=0 to 5) for \finalnumberofsources\
extragalactic targets observed with \FUSE. All of these have a final
signal-to-noise ratio larger than \snlimit, and are located at galactic latitude
\babs$>$20\deg. The individual observations were calibrated with the \FUSE\
calibration pipeline CalFUSE version 2.1 or higher, and then carefully aligned
in velocity. The final velocity shifts for all the \FUSE\ segments are listed.
\H2\ column densities or limits are determined for the 6 lowest rotational ($J$)
levels for each \HI\ component in the line of sight, using a curve-of-growth
approach at low column densities ($\ltsim$16.5), and Voigt-profile fitting at
higher column densities. Detections include \finaldetections\ measurements of
low-velocity \H2\ in the Galactic Disk and lower Halo. Eight sightlines yield
non-detections for Galactic \H2. The measured column densities range from
log\,$N$(\H2)=14 to log\,$N$(\H2)=20. Strong correlations are found between
log\,$N$(\H2) and \T01, the excitation temperature of the \H2, as well as
between log\,$N$(\H2) and the level population ratios
(log\,($N(J^\prime)/N(J)$)). The average fraction of nuclei in molecular
hydrogen ($f$(\H2)) in each sightline is calculated; however, because there are
many \HI\ clouds in each sightline, the physics of the transition from \HI\ to
\H2\ can not be studied. Detections also include \H2\ in  \ivdetections\
intermediate-velocity clouds in the Galactic Halo (out of \ivlos\ IVCs).
Molecular hydrogen is seen in one high-velocity cloud (the Leading Arm of the
Magellanic Stream), although \hvlos\ high-velocity clouds are intersected; this
strongly suggests that dust is rare or absent in these objects. Finally, there
are five detections of \H2\ in external galaxies.
\end{abstract}


%%%%%%%%%%%%%%%%%%%%%%%%%%%%%%%%%%%%%%%%%%%%%%%%%%%%%%%%%%%%%%%%%%%%%%%%%%%%%%%%
%%%%%%%%%%%%%%%%%%%%%%%%%%%%%%%%%%%%%%%%%%%%%%%%%%%%%%%%%%%%%%%%%%%%%%%%%%%%%%%%

\section{Introduction}
Molecular hydrogen (\H2) is an important constituent of the interstellar medium
(ISM), especially in the denser parts, where the presence of \H2\ is connected
with star formation. In its ground electronic state \H2\ has many absorption
lines in the ultraviolet part of the spectrum, between about 900 and 1130~\AA.
Using the {\it Copernicus} satellite, \H2\ was detected in many sightlines
toward nearby disk stars (Spitzer \& Jenkins 1975, Savage et al.\ 1977), showing
column densities between \dex{14} and \dex{20}~\cmm2\ for sightlines with N(\HI)
below $\sim$5\tdex{21}~\cmm2. At higher values of N(\HI) a large fraction of the
hydrogen becomes molecular. The formation of \H2\ on interstellar dust particles
is theoretically understood (Hollenbach et al.\ 1971). The transition from
mostly neutral to mostly molecular to gas was worked out by Federman et al.\
(1979).
\par After {\it Copernicus} there was a long hiatus in the possibility of
measuring interstellar \H2. The {\it IMAPS} (Jenkins et al.\ 1988) and {\it
ORFEUS} (Kr\"amer et al.\ 1990) payloads on the {\it ASTRO-SPAS} space-shuttle
platform provided R=\dex5 and \dex4 resolution data in the 900--1100\,\AA\
spectral range, although for only a limited number of targets with V magnitudes
$\ltsim$12. The launch of the {\it Far Ultraviolet Spectroscopic Explorer}
(\FUSE) in 1999 resulted in much better access to the wavelength region
containing the \H2\ lines. The properties of the \FUSE\ satellite have been
described in detail by Moos et al.\ (2000) and Sahnow et al.\ (2000). The
sensitivity of \FUSE\ allows observing extragalactic and stellar background
targets with fluxes down to about \dex{-14}~\fu\ (corresponding to a V magnitude
of about 16).
\par The molecular hydrogen lines in the FUV are associated with transitions
from the ground electronic state, X$^1$$\Sigma_g^+$ to excited electronic
states, B$^1$$\Sigma_u^+$ (Lyman bands) and C$^1$$\Pi_u$ (Werner bands). The
transitions start from the ground ($v$=0) vibrational state and a range of
rotational states, $J$. In practice transitions starting with $J$=0 to $J$=5 are
observed. The even states ($J$=0, 2, 4, ...) have a nuclear spin $S$=0
(para-\H2), while the odd states ($J$=1, 3, 5,...) have nuclear spin $S$=1
(ortho-\H2). Abgrall et al.\ (1993a, b) presented detailed tables of the
wavenumbers and Einstein-A values for all these transitions. However, these are
out of date, and this paper instead use the oscillator strengths, wavelengths
and damping constants provided by Abgrall to the FUSE PI Team (courtesy of Ken
Sembach).
\par The \H2\ molecule provides some diagnostics of the physical conditions in
the interstellar medium (see e.g.\ review by Shull \& Beckwith 1982). The
population ratios of the different rotational states can be described by a
Boltzmann exponential ($N(J)\propto \exp\left(-E/kT\right)$). In practice, the
Boltzmann temperatures are often just a convenient shorthand for describing the
rotational level population ratios. However, the lowest of these temperatures
(\T01) does provide a measure of the collisional excitation. In diffuse clouds,
the higher rotational states are also excited by FUV radiation, and the
populations of the higher $J$ levels are governed by radiative excitation,
followed by UV fluorescence, infrared radiative cascades and collisional
de-excitation. These processes are described in several papers that discuss
modeling of the population levels of \H2\ in the diffuse ISM (see Browning et
al.\ 2003 and many references therein, most usefully Spitzer \& Zweibel 1974,
Jura 1975a,~b). In somewhat denser clouds the higher $J$-levels may also be
collisionally excited, for which Gry et al.\ (2002) present some evidence.
\par Toward almost all Galactic and extragalactic targets observed with \FUSE\
Galactic low-velocity \H2\ absorption is detected. Published studies include
surveys of \H2\ in diffuse clouds in the Milky Way, using disk OB stars
(Rachford et al.\ 2002), in interstellar gas toward selected Seyferts and
quasars (Sembach et al.\ 2001a, b), intermediate- velocity clouds (Richter et
al.\ 2003), high-velocity clouds (Richter et al.\ 2001), the Magellanic Clouds
(Tumlinson et al.\ 2003) and M\,33 (Bluhm et al.\ 2003).
\par As of 31 Dec 2004, \FUSE\ has observed almost 400 extragalactic background
targets (excluding LMC/SMC stars in this count). For 201 of these the
signal-to-noise ratio (S/N) per 20~\kms\ resolution element near 1030\,\AA\ is
$>$3. However, for datasets with S/N$<$\snlimit, the \H2\ becomes difficult to
measure, and therefore this paper only includes the \finalnumberofsources\
targets with S/N$>$\snlimit. These targets are located at high galactic
latitudes (\babs$>$20\deg), and probe the diffuse \H2\ in the Galactic Halo and
the Disk near the Sun. They were originally observed for many different
purposes, including the original survey of AGN sightlines by the \FUSE\ Team
aimed at measuring Galactic \OVI\ absorption, high-velocity clouds and Galactic
deuterium. Other targets come from general observer programs in \FUSE\ Cycles 1
through 5, which include studies as diverse as searching for intergalactic \OVI,
measuring metals in Galactic high-velocity clouds and analyzing properties of
starburst galaxies. The \H2\ in these sightlines is often a contaminant for
accomplishing the original aim of the observation. In fact, the current survey
was started as a way to understand which features of interest might instead be
\H2\ and to be able to remove the \H2\ lines. This paper therefore will
concentrate on presenting the data analysis and the \H2\ measurements, and only
a minimal attempt is made at interpreting the results and discussing the
implications for understanding the physical conditions of the Galactic ISM.
\par In \Sect\Sobserve\ the observations are discussed. Section~\Smeasure\
presents the methods used to determine N(\H2), while \Sect\Sresults\ describes
the results.

%%%%%%%%%%%%%%%%%%%%%%%%%%%%%%%%%%%%%%%%%%%%%%%%%%%%%%%%%%%%%%%%%%%%%%%%%%%%%%%%
%%%%%%%%%%%%%%%%%%%%%%%%%%%%%%%%%%%%%%%%%%%%%%%%%%%%%%%%%%%%%%%%%%%%%%%%%%%%%%%%

\section{Observations and data reduction}
\subsection{The sightline sample}
\par The sightlines in the sample are listed in Table~\Tobservations. Columns~1,
2, 3 and 4 give the object name, galactic longitude and latitude, and the target
type (QSO, Seyfert or Hubble type). Columns~5, 6 and 7 list the \FUSE\ datasets,
the version of the calibration pipeline that was used, and the exposure time.
Many targets were observed multiple times. All these exposures and segments were
combined before measuring the \H2\ lines, in the manner discussed in
\Sect\Sshift\ below. Each individual observation was obtained using the \FUSE\
time-tag mode, almost always through the 30\arcsec$\times$30\arcsec\ LWRS
aperture. A note ``3'' to Col.~5 in Table~\Tobservations\ indicates the few
datasets obtained through the 4\arcsec$\times$20\arcsec\ MDRS aperture.
\par Each observation provides a photon list for each of a set of consecutive
exposures corresponding to an orbital viewing period. The photon lists for each
exposure are concatenated before being processed by the \FUSE\ calibration
pipeline. Wakker et al.\ (2003) provide a discussion of the steps involved in
the calibration. However, they used the CalFUSE v1.8.7 version of the pipeline,
which had some known problems, most notably an inaccurate wavelength scale.
Later CalFUSE versions corrected this problem and also have better flux
calibration, better modeling of event bursts, scattered light and astigmatism,
error propagation, jitter correction and bad-pixel rejection. Therefore, we
started recalibrating our old datasets with CalFUSE v2.1 as soon as it became
available. When CalFUSE v2.4 became available, we switched to that version. In
principle all datasets should have been recalibrated with CalFUSE v2.4, but due
to lack of manpower and time about half are still v2.1. However, for the few
cases where both v2.1 and v2.4 calibrations exist, there are no obvious
differences between the two datasets.
\par The exposure times listed in Col.~7 of Table~\Tobservations\ are the
nominal times listed in the \FUSE\ observation log. Because of ``bursts''
(short-lasting large increases in the detector count rate) and passages through
the South Atlantic Anomaly (SAA) as well as other such problems, part of the
exposure may be unuseable. If the actual exposure time on either Side 1 or Side
2 of the detector is more than 10\% less than the nominal time, the actual time
is given in parentheses, with a note ``1'' or ``2'' indicating whether the loss
of time pertains to Side 1 or Side 2.

%%%%%%%%%%%%%%%%%%%%%%%%%%%%%%%%%%%%%%%%%%%%%%%%%%%%%%%%%%%%%%%%%%%%%%%%%%%%%%%%

\subsection{Wavelength shifts}
\par A \FUSE\ spectrum consists of data for 8 detector segments, covering
different, partially overlapping wavelength regions: LiF1A (987--1082~\AA),
LiF1B (1094--1187~\AA), LiF2A (1085--1181~\AA), LiF2B (983--1074~\AA), SiC1A
(1004--1091~\AA), SiC1B (905--993~\AA), SiC2A (917--1006~\AA), SiC2B
(1017--1105~\AA). The sensitivity of the LiF1A and LiF2B segments is so much
higher than that of SiC1A and SiC2B that the latter are rarely used. Similarly,
the SiC2A data are of much better quality than those for SiC1B, so the SiC1B
channel is ignored here. However, in the wavelength regions where they overlap,
the S/N ratio provided by LiF2B/LiF1B is about half that of LiF1A/LiF2A, and it
usually makes sense to combine the data for those segments, even though they
have slightly different spectral resolutions. The SiC2B data cover a wavelength
range similar to that of LiF1A and LiF2B, but they are much noisier and are only
used in the wavelength region where they provide unique data (1082 to 1085~\AA).
\par \FUSE\ wavelengths are calibrated using a wavelength solution derived from
a number of bright targets with well-studied and simple interstellar features
(Sahnow et al.\ 2000). However, problems like minor misalignments in target
positioning lead to zero point offsets that are typically less than 10~\kms\
(half a resolution element, or 5 detector pixels), but which can be as large as
20~\kms. These offsets vary between different targets, sometimes between
different observations of the same target taken at different times of the year,
and also between different segments in the same observation. Offsets often occur
for slightly extended sources, whose UV emission may peak at a different
location than the optical position used to point the telescope. Correctly
correcting for these offsets involves a major effort, without which absorption
lines can be smeared and different segments and different exposures of the same
target may end up misaligned.
\par Wakker et al.\ (2003) found shifts for the LiF1A and LiF2B segments by
measuring the centroids of the interstellar \ArI-1048 and \SiII-1020 lines,
which span their line of interest -- \OVI\l1031.926. This is insufficient for a
study of \H2, however. First, it is not a-priori true that the velocity of the
\H2\ is the same as that of the metal lines. This was demonstrated by Sembach et
al.\ (2001b) in the sightline toward 3C\,273.0, and it can also be seen by
comparing the entries in Cols.~8 and 9 of Table~\Tvelocities. Second, shifts are
needed for each detector segment, not just LiF1A/LiF2B. Third, for many sources
the S/N ratio in each individual exposure is low and the centroids of the
\ArI-1048 and \SiII-1020 lines are difficult to determine.
\par A more rigorous approach is therefore taken in this paper. First, gaussians
are fitted to an unshifted spectrum to find the centroids of all discernable
metal and \H2\ lines in each of the segments of each individual observation. In
a number of sightlines intermediate- and/or high-velocity metal line absorption
is seen, in which case separate centroid fits are made to these components. The
metal lines usually include \SiII\l1030.699, \ArI\l1048.220, \ArI\l1066.660,
\FeII\l1063.177 for the LiF1A/LiF2B segments and \FeII\ll1144.938, 1121.975,
1143.226, 1125.448, \PII\l1152.818 for the LiF1B/LiF2A segments. In many (but
not all) sightlines, \FeII\ll1081.875, 1096.877 are also included (if not
contaminated by \H2), and sometimes \FeII\ll1055.262, 1062.152, 1063.972,
1112.048, 1127.098, 1133.665, 1142.366 (if strong enough). Occasionally
\OI\l1039.230 and/or \NI\ll1134.165, 1134.415, 1134.980 are not contaminated by
geocoronal emission and can be used. Metal lines uncontaminated by geocoronal
emission are rare in the SiC2A segment, but selecting only the data taken during
orbital night may allow measurements of \OI\ll948.686, 936.630, 976.448,
929.517, 950.885, 924.952, \NI\ll953.970, 953.655, 963.990, 953.415, 964.626,
965.041, and \PII\l963.081. Depending on the strength of the \H2\ lines,
anywhere between 0 and 40 \H2\ lines can be discerned in any of the six analyzed
segments. For sightlines with log\,$N$(\H2)=14--15 several lines of $J$=0 and
$J$=1 are strong and narrow enough to give reliable centroids. For
log\,$N$(\H2)$\sim$15--17 lines for $J$=2 and $J$=3 and the weaker $J$=0 and
$J$=1 lines are the most reliable. At higher $N$(\H2) most of the $J$=0 and
$J$=1 develop damping wings, and a centroid is no longer fitted. Lines that are
blended with other \H2\ or metal lines are not fitted either. The fitted
centroids are then plotted against wavelength and separate average velocities
are determined for the metal lines and the \H2\ lines. The rms spread in the
velocities ranges from about 2~\kms\ for sightlines with high signal-to-noise
ratio and simple absorption-line structure, to 6--8~\kms\ for low
signal-to-noise data.
\par For seven of the ten datasets in the sample that were obtained before 2
November 1999 (as well as for three later ones) a clear linear slope can be seen
in the centroid vs wavelength plot for some (usually not all) segments, even
after applying the v2.1.6 or v2.4.0 calibration. This slope is typically about
0.11 to 0.15~\kms/\AA, giving a total centroid shift of $\sim$20--30~\kms\
across a segment. Not taking this into account would smear out \H2\ lines at the
edges of the segment and lead to erroneous equivalent width measurements. The
values for these special slopes can be found in Note~4 to Table~\Tobservations.
\par Two methods were used to determine the expected velocity of the metal
lines. Twenty-one sightlines have data taken with the E140M grating on the Space
Telescope Imaging Spectrograph (\STIS). This spectrograph has better velocity
resolution than \FUSE\ (6.5~\kms\ vs 20~\kms), and its wavelength calibration
appears to be more reliable, due to the much smaller slit (0\farcs2) (see
appendix in Tripp et al.\ 2005). STIS data are provided on a heliocentric
wavelength scale, and a heliocentric to LSR correction needs to be applied.
Centroids were fitted to the metal lines in the STIS wavelength range that show
clean profiles, which can include \NI\ll1199.550, 1200.223, 1200.710,
\SII\ll1250.584, 1253.811, 1259.519, \NiII\ll1317.217, 1370.132, 1454.842,
\FeII\l1608.451, and \AlII\l1670.787. In general, they show an rms spread of
about 1--2~\kms. The centroids of the \FUSE\ metal lines are then aligned with
the velocity implied by the STIS-E140M data.
\par If no STIS E140M data are available, the velocity of the metal lines is
compared to that of the 21-cm \HI\ emission. If there are well-detected
intermediate- or high-velocity components, these are used preferentially. The
\HI\ spectrum can come from one of three possible sources, as described by
Wakker et al.\ (2003), who show these \HI\ spectra: a) special observations with
the Effelsberg 100-m dish for a limited number of sightlines, b) the
Leiden-Dwingeloo Survey (LDS, Hartmann \& Burton 1997), or c) the Villa Elisa or
IAR Survey (Arnal et al.\ 2000). The assumption that the centroid of the metal
lines aligns with the peak of the \HI\ emission often turns out to be incorrect.
For the 19 sightlines that have STIS-E140M data a comparison of $<$$v$$>$(STIS)
and $<$$v$$>$(21-cm) yields a range of $\pm$10~\kms\ for the difference in
velocities, with an average difference of 2~\kms\ and an rms of 4~\kms. However,
even if the absolute velocity of the metal lines will not be exactly right, the
relative alignment of the different FUSE detector segments will be.
\par There are five sightlines (1H\,0707$-$495, Mrk\,1095, MS\,0700.7+6338,
NGC\,1068 and PG\,0844+349, see Col.~10 of Table~\Tvelocities), where the \HI\
spectrum shows a very narrow, very bright low-velocity component superimposed on
a broader profile. In these cases the velocity of the \H2\ (rather than that of
the metal lines) was aligned with this probably cold \HI.
\par The implied shifts usually show distinct patterns. The shifts implied by
the metal and \H2\ lines are often similar for all LiF1 and LiF2 segments, or
for all segments on Side~1 and Side~2, or even for all LiF segments. Shifts that
were similar to within about 1$\sigma$ were averaged to find the final implied
shifts for those segments. Shifts that are clearly different from each other
were averaged separately. The shifts implied for SiC2A and SiC2B are often very
different from those implied for the LiF segments. For exposures with low S/N,
the SiC2B data are often too noisy to discern the absorption lines, and the
SiC2A shift is assumed.
\par After determining the shifts for all segments, the LiF1A+LiF2B and
LiF2A+LiF1B data from different observations are combined. Then the centroids of
the metal and \H2\ lines are fitted again, to check whether the average centroid
velocities are now indeed equal to those implied by the STIS or the 21-cm data.
Another iteration is done if they are not. The final shifts for each segment are
presented in Cols.~8 to 13 of Table~\Tobservations. For several observations the
LiF2B or LiF1B segment shows a much lower flux than LiF1A or LiF2A, and is
ignored when combining the segments; in other cases one or more segments do not
show any flux. The unused segments are indicated by a ``$-$'' in
Table~\Tobservations.
\par Column.~2 to 6 of Table~\Tvelocities\ present fluxes and S/N ratios for
each target, determined near four wavelengths: 1031, 1063, 977 and 1122\,\AA.
These are representative of the LiF1A+LiF2B, LiF1A+LiF2B, SiC2A, and LiF2A+LiF1B
data, respectively. Two values are given for the LiF1A+LiF2B data, since in some
sightlines there is redshifted Lyman-$\beta$ intrinsic to the background galaxy
that lowers the flux and S/N ratio near 1031\,\AA\ compared to the overall
values in the segment. This is also reflected in the flux value given in Col.~2,
which is usually valid near 1031\,\AA, but is for near 1063\,\AA\ if a note
($^a$) is given. Measuring the S/N ratios is done by fitting a 1st, 2nd or 3rd
order polynomial to the continuum within about $\pm$1500~\kms\ of the selected
wavelength, and then calculating the rms of the residual of the fit. This is
compared with the error array given by the calibration software to make sure
that the rms makes sense -- the rms of the residual should not be smaller than
the errors; it generally is not more than 20\% larger. For a few high
signal-to-noise sightlines (3C\,273.0, Mrk\,509, NGC\,1705, NGC\,4151, 
PKS2155$-$304) fixed-pattern noise limits the S/N ratio to slightly better than
30. For other high S/N sightlines (Mrk\,279, Mrk\,421, Mrk\,817, Mrk\,876,
PG\,0804+761, PG\,1259+593) there are multiple slightly-differently-aligned
exposures, reducing the fixed-pattern noise, and allowing and S/N$>$30.
\par The \HI, metal-line and \H2\ velocities that were measured are given in
Cols.~7 to 9 of Table~\Tvelocities. Column~7 lists the velocities of the \HI\
components found in the 21-cm spectra, which can be found in Wakker et al.\
(2003). Column~8 shows the metal-line velocities that can be derived from the
alignment method described above, while Col.~9 gives the implied \H2\ velocity.
The distribution of $v$(metal)$-$$v$(\H2) has an average of 0.4~\kms, and a
dispersion of 3.9~\kms. The largest differences ($\pm$9~\kms) occur for
PG\,1259+593 and PG\,1302$-$102. Column~10 shows how the different kinds of data
were aligned. ``E140M'' means that there is \STIS\ data from which $v$(metal)
follows, and the \FUSE\ and \STIS\ metal lines can be aligned, implying
$v$(\H2). In this case $v$(\HI) can be different from $v$(metal). ``21cm'' in
Col.~10 means that the \FUSE\ metal lines were aligned with the 21-cm data.
Finally, ``21cm-H2'' implies that it was assumed that the \H2\ aligns with a
bright, narrow component in the 21-cm \HI\ spectrum.

%%%%%%%%%%%%%%%%%%%%%%%%%%%%%%%%%%%%%%%%%%%%%%%%%%%%%%%%%%%%%%%%%%%%%%%%%%%%%%%%
%%%%%%%%%%%%%%%%%%%%%%%%%%%%%%%%%%%%%%%%%%%%%%%%%%%%%%%%%%%%%%%%%%%%%%%%%%%%%%%%

\section{Determination of \H2\ parameters}
\par After properly aligning and combining the individual observations and
segments, the \H2\ column densities for each rotational ($J$) level can be
measured. This is done in several steps. 1) A piecewise global continuum is fit
to the combined data. 2) For each $J$ level the continuum-normalized absorption
lines are displayed all at once, and lines that are not blended, confused, or
too noisy are selected interactively. 3) For log\,$N(J$)$\ltsim$15.8, a
curve-of-growth is used to determine the column density and estimate the
$b$-value, while for log\,$N(J$)$\gtsim$16.6 a Voigt profile is fit to the
damping wings; for log\,$N(J$) between these two values either method may be
used. 4) The column densities of the up to five different $J$ levels are
compared and parametrized with five parameters: $N$(\H2,total), $b$, \T01, \T23,
and \T34. 5) The parametrization can be turned into a theoretical absorption
profile for each line, which is overlaid on the data, in order to make sure that
it makes sense. Now follow comments on each of these steps.
\medskip
\par {\it 1) Continuum fitting.} In principle one could fit a local continuum to
each individual \H2\ absorption line, using e.g.\ a $\sim$1500~\kms\ wide
velocity range (with all continua spliced together). However, in practice this
is prohibitively time-consuming. Therefore, 1st to 4th order Legendre
polynomials (see Sembach \& Savage 1992) were fitted to line-free regions over
up to 30\,\AA\ wide wavelength ranges.
%over the following wavelength ranges: 920 to 950~\AA, 950 to 975~\AA, 975 to
%1000~\AA\ (using SiC2A data), 1000 to 1022~\AA, 1022 to 1025~\AA\ (i.e.\ in the
%Milky Way Ly$\beta$ damping wing), 1028 to 1050~\AA, 1050 to 1075~\AA\ (the
%latter four regions use the combined LiF1A+LiF2B data), 1075 to 1082~\AA\ (LiF1A
%only), 1090 to 1095~\AA\ (LiF2A only), 1095 to 1125~\AA\ (LiF2A+LiF1B). However,
%in several sightlines intrinsic \CIII, \OVI, or Ly$\beta$ emission yields bright
%features in the spectra. In those cases the wavelength ranges above were
%adapted to allow proper fits across these emission features.
\medskip
\par {\it 2) Line selection.} After normalizing, the lines for each $J$ level
are displayed (sorted by oscillator strength), and lines for which reliable
equivalent widths can be derived are selected interactively. Lines that are
always blended with interstellar metal lines are excluded, as are lines blended
with other \H2\ lines. Also excluded are lines that are blended with
intergalactic absorption, intrinsic AGN lines, or geocoronal emission. For
sightlines with S/N(SiC2A) less than about 8, all lines with
$\lambda$$<$1000~\AA\ are usually excluded. Displaying all lines of a single
$J$-level at once allows a check on whether the progression from strong to weak
lines conforms to expectations.
\medskip
\par {\it 3) Column density measurement.} If $N(J$)$\ltsim$\dex{16}~\cmm2, a
$\chi^2$ fit is made. First, equivalent widths are determined by direct
summation of the absorption over a $\sim$25~\kms\ velocity extent. A $\chi^2$
value is then calculated for every combination of log\,$N$ and $b$:
\begin{equation}
\chi^2\ =\ \Sigma\ { (\,{\rm EW}_{\rm obs}\ -\ {\rm EW}_{\rm pred}(N,b)\,)^2 \over \sigma(EW)^2 }.
\end{equation}
Minimizing $\chi^2$ gives a ``best-fit'' column density and $b$-value for each
individual $J$-level. 
\par Of course, the resulting column density strongly depends on the $b$-value.
Due to noise, uncertainties in continuum placements, and the relatively small
range in $f\lambda$ values for each $J$-level the best-fit $b$-value generally
differs for the different $J$-levels. Further, at low column densities the
equivalent widths are independent of $b$, while at intermediate column densities
the derived value of $N$ is very sensitive to $b$. Therefore, it is assumed that
$b$ is the same for each $J$-level, and the resulting column density is the one
that gives the lowest value of $\chi^2$, given a fixed value of $b$. How $b$ is
determined is described in point 5) below.
\par A few papers have claimed evidence for an increase in $b$ with $J$ (Spitzer
\& Cochran 1973, Spitzer et al.\ 1974, Lacour et al.\ 2005). These authors
determined curves-of-growth toward high signal-to-noise Copernicus and \FUSE\
spectra of nearby disk stars with high \H2\ column densities. Only a few of the
spectra of the extragalactic targets in this paper have comparable S/N ratios,
and they all show much lower \H2\ column densities. Limiting the check to the 18
sightlines with S/N$>$25, reliable independent $b$ values for multiple $J$
levels can be determined for just two \H2\ components: $J$=1 to 3 at +28~\kms\
toward 3C\,273.0 and $J$=0 to 2 in the IVC toward PG\,1116+215. The following
FWHMs are found for 3C\,273.0; FWHM($J$=1)=8.0$\pm$1.1, FWHM($J$=2)=7.7$\pm$1.3,
FWHM($J$=3)=7.0$\pm$1.4; for PG\,1116+215: FWHM($J$=0)=6.6$\pm$1.4,
FWHM($J$=1)=7.0$\pm$1.4, FWHM($J$=2)=5.0$\pm$2.5. Thus, these two sightlines do
not show clear evidence for systematic variations of the FWHM with $J$.
\par The determination of column densities is complicated by the fact that the
$f\lambda$ values of the \H2\ lines of a given level span only a relatively
narrow range. For $J$=0, and $\lambda$$>$1000~\AA, just seven lines are
unblended, and the five strongest differ by just a factor 2.3 in $f\lambda$; the
two most important lines are therefore L(1-0) R(0) $\lambda$1092.195, and L(0-0)
R(0) $\lambda$1108.127, which are 4.8 and 16.6 times weaker than the strongest
$J$=0 line. For $J$=1 the situation is slightly better, with 13 useful lines
spanning a factor 6 in $f\lambda$, and two lines whose strength is 1/12.3 and
1/20.3 that of the strongest lines (L(1-0) P(1) $\lambda$1094.052, L(0-0) R(1)
$\lambda$1108.633). For $J$=2 15 lines span a factor 6.5 in $f\lambda$, with one
more (L(1-0) P(2) $\lambda$1096.44) 1/10.4 times the strength of the strongest.
Thirteen $J$=3 lines span a factor 2.5 in $f\lambda$, with two substantially
weaker lines (L(1-0) P(3) $\lambda$1099.787 and L(0-0) P(3) $\lambda$1115.895)
factors of 7.6 and 26 below the strongest line. If $J$=4 or $J$=5 is detected,
even the strongest lines are on the linear part of the curve-of-growth, so that
there are no problems in measuring $N$(4) and $N$(5). The weakest lines of each
$J$ level are often crucial in determining the column density of a given
$J$-level. Only in sightlines where log\,$N(J)$$\ltsim$15 do the stronger \H2\
lines contain significant column density information. On the other hand, the
stronger lines allow one to obtain a reasonable estimate of $b$, which is set by
the maximum equivalent width.
\par At high column densities ($N(J)$$\gtsim$\dex{17.5}~\cmm2), some or all of
the \H2\ $J$=0--3 absorption lines develop damping wings. In these cases a Voigt
profile was fit to the damping wings, leading to relatively reliable column
densities. In the column density range log\,$N(J)$=15.8--16.6, which method was
used depends on the implied $b$-value, and on the ease of fitting a Voigt
profile to the strongest lines. The decision is sightline dependent -- details
can be provided upon request. If there are multiple \H2\ components, the
procedure outlined above was repeated for each component separately. The
selection of lines used may differ, depending on what blending with other lines
is present. In many cases no lines of a particular $J$ level are detectable, and
only an upper limit can be derived. The numerical value of of this limit is set
to the column density corresponding to twice the error on the equivalent width
of the strongest unblended line(s), assuming it has an FWHM of 9~\kms. Two sigma
is used because there are always several lines that give similar limits and that
thus reinforce each other. This is also obvious from overplotting the
theoretical profiles.
\medskip
\par {\it 4) Obtaining temperatures.} 
Given a $b$-value (see point 5 below), the method described above yields a set
of column densities. The ratios of $N(J+1)$ to $N(J)$ can be described using the
Boltzmann distribution:
\begin{equation}
{N(J+1) \over N(J)} = {g_{J+1} \over g_J}\ \exp\left({-(E_{J+1}-E_J) \over k T_{J(J+1)}}\right),
\end{equation}
with $g_J$ the statistical weights (1, 9, 5, 21, 9, 33 for $J$=0, 1, 2, 3, 4, 5)
and $E_J$ the excitation energies of the different levels (0, 0.01469, 0.04394,
0.08747, 0.14491 and 0.21575~eV, respectively) (see Abgrall et al.\ 1993a).
Thus, the five column density ratios $N$(1)/$N$(0), $N$(2)/$N$(1),
$N$(3)/$N$(2), $N$(4)/$N$(3), and $N$(5)/$N$(4) can be converted to five ``ratio
temperatures'', \T01, \T12, \T23, \T34, T45. If the excitation of the \H2\
molecule is purely collisional and in equilibrium, these five temperatures will
be the same. In reality, $T$ increases for higher level ratios because of
radiative excitation. Here, the ratio temperatures are considered purely as way
to parametrize the column densities: the (up to) six measured column densities
can be parametrized by (up to) six parameters: four ratio temperatures, the
total \H2\ column density and the intrinsic width.
\medskip
\par {\it 5) Deriving $b$.} At this point the derivation of the preferred value
of $b$ can be discussed. There are two methods, which give comparable results
when they can both be applied. First, one can do a combined $\chi^2$ fit to the
equivalent widths of all lines without damping wings to derive a best-fit $b$.
However, it turns out that this method is not always robust. Sometimes the
resulting value for $b$ implies wildly varying ratio temperatures (including
\T23$<$\T01, which is not physical). At high $N$(\H2) often just a few $J$=4 or
$J$=2/3 lines can be used. At low $N$(\H2) most lines are on the linear part of
the curve-of-growth and thus the column densities are insensitive to $b$. The
best case happens at intermediate $N$(\H2) where reasonable curves-of-growth can
be made. However, except for two sightlines (see below) the data are too noisy
to provide good constraints on $b$ based on more than two $J$-levels.
\par A second method, used for the remainder of this paper, is to assume that
\T12=\T23. Then there is only one value of $b$ at which the value of $N$(2)
predicted from \T12 (=\T23) and $N$(1) is the same as the observed value of
$N$(2), given a value for $b$. This method is robust, and gives unique
solutions. It was invented after the results obtained for temperatures and
linewidths using more traditional methods often yielded \T12$\sim$\T23.
Therefore, it was decided to enforce this equality for sightlines where the data
do not readily allow one to derive \T12\ and \T23\ independently. It should be
noted, however, that the assumption that \T12=\T23 may be incorrect, especially
for sightlines through denser disk gas. However, only for sightlines that have
high signal-to-noise data combined with intermediate \H2\ column density
(log\,$N$$\sim$15.3 to 16.3) is it possible to measure independent $b$ values
for several $J$ levels. For the current sample of high-latitude extragalactic
targets with S/N$>$\snlimit\ \FUSE\ data the assumption that \T12=\T23 can be
justified by the following considerations.
\par 1) One justification comes from the 30 \H2\ components where a combined
$\chi^2$ fit to $b$ does yield a useful answer because there are many weak and
strong $J$=2, 3 and/or $J$=4 lines. The $b$-values from the \T12=\T23 method
are given in Col.~9 of Table~\Tmolhyd, those from a direct fit in Col.~10. The
correlation coefficient of $b$($\chi^2$ fit) vs $b$(\T12=\T23) is 0.85. The
ratio of the two has a distribution with average 1.00 and dispersion 0.10, while
their difference is $-$0.1$\pm$0.9~\kms. Note that the typical uncertainty on
$b$ is about 2~\kms. So, in the cases where $b$ could be estimated in two ways
the two methods give consistent results.
\par 2) A second justification comes from the two components with good S/N and
intermediate-strength \H2\ where $b$ can be fitted independently for three $J$
levels: at +28~\kms\ toward 3C\,273.0 and at $-$44~\kms\ toward PG\,1116+215.
Using the FWHMs and column densities found separately for each $J$, these yield
\T01=209$\pm$30~K, \T12=295$\pm$25~K, \T23=285$\pm$15~K for 3C\,273.0 and
\T01=163$\pm$32~K, \T12=222$\pm$36~K, \T23=219$\pm$22~K for PG\,1116+215~K.
Compare these to the values in Table~\Tmolhyd: \T01=184~K, \T23=270~K for
3C\,273.0, and \T01=167~K, \T23=219~K for PG\,1116+215. It is clear that \T12
equals \T23 to within the errors, but also that it is different from \T01.
\par 3) A final justification can found by trying to make fits assuming that
\T12=\T01, rather than \T12=\T23. However, not all sightlines are suitable for
this comparison. First, for sightlines with $N$(\H2)$\gtsim$18, the $J$=0 and
$J$=1 lines have damping wings, and thus $N$(0), $N$(1) and \T01 are independent
of $b$. Also, as can be seen from Table~\Tmolhyd, \T01 and \T23 are rather
similar for these sightlines, so \T12 will be similar to either one. Second,
since $N$(\H2) is more difficult to fit for sightlines with two-component \H2,
these are not suitable for a detailed test. Third, targets with extended UV
emission are also unsuitable. Finally, a high S/N ratio is best. There are 15
targets with S/N$>$15, log\,$N$(\H2)$<$16, and a single \H2\ component:
Mrk\,279, Mrk\,817, PG\,1259+593, PKS\,2155$-$303, 3C\,273.0, Mrk\,421,
H\,1821+643, HE\,0226$-$4110, PG\,0953+414, Mrk\,1383, NGC\,4670, Ton\,S210,
PKS\,0558$-$504, PKS\,0404$-$12, PKS\,2005$-$489. For all these one combination
of $N$(\H2,total), $b$, \T01, \T23 and \T12=\T23 yields a good parametrization
of the observed $N$(0) to $N$(3). However, assuming \T12=\T01 always leads to a
value for $N$(2) predicted from $N$(1) and \T01 that lies a factor 2 to 3 below
the observed value. That means that there is no value for $b$ that gives a
combination of $N$(0) to $N$(3) for which the ratio $N$(1)/$N$(0) implies the
same temperature as the ratio $N$(2)/$N$(1), but there is a value of $b$ for
which the $N$(3)/$N$(2) ratio implies a temperature that also fits the ratio
$N$(2)/$N$(1).
\par In summary, it is always possible to parametrize the five observed column
densities with five observed parameters log\,$N$(\H2,total), $b$, \T01, and
\T23, \T34, with \T12=\T23. This last assumption was tested in a number of
sightlines, and is applied to all sightlines.
\medskip
\par Determining the column density is difficult, but obtaining useful errors
adds even more complications. There are several sources of uncertainty. The
noise in the data and the placement of the continuum allow the calculation of an
error for each individual equivalent width measurement. In principle, estimating
the continuum placement error requires fitting a separate continuum over a small
wavelength range near each absorption line. Since this is impractical because of
the large number of targets in the sample, a larger wavelength range was used to
fit the continuum across many \H2\ lines at once. This results in
underestimating the contribution of the continuum placement error to the
equivalent width error. However, the targets all have relatively high S/N and
mostly flat continua, so the continuum placement error is always small relative
to the error associated with the noise itself.
\par Given a set of equivalent widths and errors, the error in $N$(\H2) is found
as follows. First the best-fit value of $b$ is used to estimate errors in
$N$(J), using the 68\% confidence interval in $\chi^2$. Then an estimate for the
error in $b$ follows from the minimum and maximum value of $b$ for which the fit
for $N$(\H2), $b$, \T01 and \T23\ predicts values for $N$(2) and $N$(3) that are
still within the 1~$\sigma$ error of those column densities. Next, the errors in
$N$(J) are recalculated, using the implied range of $b$ values.
\medskip
\par {\it 6) Theoretical profiles.} Given the derived values for $N$(\H2), $b$,
\T01, \T23, and \T34, it becomes possible to create theoretical profiles for the
\H2\ absorption spectrum. This is done by creating a combined optical depth
profile, turning this into an absorption spectrum and then smoothing with the
instrumental profile (assumed to be gaussian). These profiles are overplotted on
the \H2\ lines, and compared to the observed profiles. The whole procedure
described in this subsection is then iterated upon until the theoretical
profiles look reasonable. In the course of this comparison it was found that the
resolution of the \FUSE\ LiF (SiC) spectra is indeed $\sim$20 (25)~\kms\ for
most observations. However, in two circumstances it was necessary to degrade it
to $\sim$25 (30)~\kms. This is the case for 27 targets with a single integration
longer than 40~ks, and for 5 more targets for which three or more integrations
had to be combined. Three targets fall in both categories, while for 8 targets
with long exposures the resolution did not appear to be degraded.
\par A few targets have extended UV emission. Thus, light from different parts
of the source is focused on a different part of the detector, effectively
increasing the width of the interstellar absorption lines. In these sightlines
it was therefore necessary to smooth the theoretical profiles with broader
gaussians in order to properly match the strong and weak lines simultaneously.
In a few cases this effect is mild, and a 30~\kms\ wide gaussian is needed
(Mrk\,116, NGC\,604, NGC\,625, NGC\,5253, NGC\,7714). In three cases the
broadness of the profiles is easily visible in the spectra themselves, and
gaussian fits to the \H2\ lines yield average widths of 45~\kms\ for NGC\,1068,
NGC\,5236 and 80~\kms\ for NGC\,3310 (see Fig.~\Fspecial e).

%%%%%%%%%%%%%%%%%%%%%%%%%%%%%%%%%%%%%%%%%%%%%%%%%%%%%%%%%%%%%%%%%%%%%%%%%%%%%%%%
%%%%%%%%%%%%%%%%%%%%%%%%%%%%%%%%%%%%%%%%%%%%%%%%%%%%%%%%%%%%%%%%%%%%%%%%%%%%%%%%

\section{Results and Discussion}

\subsection{Measurements}
\par Table~\Tmolhyd\ presents the derived \H2\ column densities for each
$J$-level, as well as the full-width-half-max ($W$) and temperatures, separately
for the low-, intermediate- and high- velocity gas visible in the 21-cm
spectrum. For some components two values of the FWHM are given, in Cols.~9 and
10. The values in Col.~10 are the ones determined from a combined
curve-of-growth fit to all \H2\ lines, while the values in Col.~9 are determined
using the \T12=\T23 assumption. Figure~\Fexamples\ shows the observed and
modeled \H2\ spectrum over a 9\,\AA\ range for twelve sightlines that span the
range of measured \H2\ column densities. This part of the spectrum (1047 to
1056\,\AA) contains lines of $J$=0, 1, 2, 3, 4 and 5, as well as two metal lines
(\ArI\l1048,220, \FeII\l1055.262). The figure shows that for
log\,$N$(\H2)$\sim$16.6 to 18.6 the apparent depth of the L(4-0) R(2)
$\lambda$1051.498, L(4-0) P(2) $\lambda$1053.284 and L(4-0) R(3)
$\lambda$1053.976 lines barely varies. This is because these lines are all
saturated, and their apparent depth is just a reflection of the smoothing with
the instrumental profile. For example, for the parameters measured for Mrk\,509
(log\,$N$(\H2)=17.31, FWHM=10.5~\kms, \T01=113, \T23=171, \T34=254):
$\tau$($\lambda$1049.367)=431,
$\tau$($\lambda$1049.960)=583,
$\tau$($\lambda$1051.033)=273,
$\tau$($\lambda$1051.498)=41,
$\tau$($\lambda$1053.284)=24,
$\tau$($\lambda$1053.976)=8.5,
$\tau$($\lambda$1047.550)=0.19.
\par In 9 sightlines one, two or three lines of $J$=5 are detected, while toward
HS\,0624+6907, Mrk\,116, and Mrk\,1095 five $J$=5 lines can be seen. The implied
values for $N$(5) are listed in the notes, as are the implied values for \T45.
The ratio of \T45\ to \T34\ is usually between 1.6 to 2.3, but there are not
enough well-determined values for \T45\ to study this systematically.
\par Low-velocity \H2\ is detected toward \finaldetections\ background targets,
with column densities ranging from 1\tdex{14}\,\cmm2\ (NGC\,1705) to
6\tdex{19}\,\cmm2\ (NGC\,7469). Upper limits are set in \finalnondetections\
directions. However, \H2\ with column density below these upper limits is seen
in a few directions, and at least some of the non-detections can be attributed
to an insufficiently high S/N ratio. Toward the sightline with the highest
$N$(\H2) (NGC\,7469) weak HD is also detected (see Wakker et al.\ 2003). HD is
not seen in any other sightline.
\par Intermediate-velocity \H2\ is detected in \ivdetections\ sightlines out of
\ivlos\ with intermediate-velocity \HI. In \Sect\SIVC\ these detections are
compared to the results of Richter et al.\ (2003), who studied 61 IVC components
in 56 sightlines, many with lower S/N ratio than the sample in this paper. They
reported 31 detections. In the current sample, high-velocity \HI\ is found
toward \hvlos\ sightlines, but only toward NGC\,3783 is high-velocity \H2\
absorption found, see \Sect\Sspecial. Finally, \H2\ is seen in five external
galaxies, as described in \Sect\Sexternal\ below.

%%%%%%%%%%%%%%%%%%%%%%%%%%%%%%%%%%%%%%%%%%%%%%%%%%%%%%%%%%%%%%%%%%%%%%%%%%%%%%%%

\subsection{Sightline discussion}
In this subsection special sightlines are discussed and comparisons are made
with previously published values.
\par Shull et al.\ (2000) listed preliminary Milky Way \H2\ column densities for
four sightlines to AGN. The results for the low-velocity component toward
Mrk\,876 are not very different: (old vs new) $N$=18.36 vs 18.13, FWHM=5 vs
7.5~\kms, \T01=135 vs 114~K, \T23=200 vs 163~K, but the intermediate-velocity
component was not reported by Shull et al.\ (2000). For ESO141$-$G55 there is
also generally good agreement, except for \T23 (19.28 vs 19.29, 10 vs 6.0~\kms,
113 vs 97~K, 112 vs 540~K. $N$(\H2) increases by a factor 2 toward PG\,0804+761
(18.78 vs 19.09, 8 vs 10.7~\kms, 113 vs 107~K, 180 vs 119~K), while the limit
toward PKS2155$-$304 was improved from 14.65 to 14.49.

\medskip
\par {\it 3C273.0.} Sembach et al.\ (2001b) measured all the intergalactic and
interstellar lines in this sightline, including the Galactic \H2. They noticed
that the \HI\ and the metal lines do not align with the \H2. A 21-cm spectrum
gives $v$(\HI,21-cm)=$-$6 and +25~\kms, while the metal lines in the STIS E140M
data are at $-$14 and +22~\kms. Sembach et al.\ (2001b) argued that
$v$(\H2)=+16~\kms, and found log\,$N$(0,1,2,3,4)= 15.00$\pm$0.30,
15.48$\pm$0.18, 14.76$\pm$0.12, 14.73$\pm$0.12, and $<$14.4, with a FWHM of
10$\pm$3~\kms, implying \T01=157$\pm$100~K. Using the method described in
\Sect\Sshift, aligning the metal lines in the FUSE data with those in the STIS
spectrum implies $v$(\H2)=+28~\kms, with $N$(0,1,2,3,4)= 15.14$\pm$0.12,
15.69$\pm$0.11, 14.90$\pm$0.07, 14.71$\pm$0.06, 13.73$\pm$0.12,
FWHM=7.2$\pm$0.2~\kms, and \T01=184~K. These column densities differ by only
about 0.1~dex ($<$1~$\sigma$), and are thus effectively the same. Sembach et
al.\ (2001b) further claimed not to find \H2\ absorption associated with the
$-$14~\kms\ metal line component. However, as Fig.~\Fspecial a shows, a weak
secondary \H2\ component with $N$(\H2)=14.84 can clearly be discerned at
$v$=$-$9~\kms. It is seen in 3 $J$=0, 14 $J$=1, 4 $J$=2 and 7 $J$=3 lines.

%\medskip
%\par{\it HE\,0226$-$4110.} Fox et al.\ (2005) list \H2\ column densities of
%$N$(0,1,2,3,4)= 13.72$\pm$0.15, 13.99$$\pm$0.23, $<$13.89, $<$13.89 and $<$13.83
%for this direction.
%PG0953+414  13.90      14.40      14.00      14.12      <14.09   14.75  Fox
%            14.13,0.15 14.75,0.16 14.11,0.18 14.13,0.18 <13.97   14.98,0.09
%            14.05,0.18 14.65,0.19 14.00,0.10 14.01,0.21 <14.12   14.82,0.13

\medskip
\par{\it Mrk\,153}. The sightline toward Mrk\,153 is extremely unusual, as in
addition to the ``normal'' Galactic \H2\ component with log\,$N$(\H2)=15.51 at
$-$20~\kms\ there is also a component with log\,$N$(\H2)=16.14 at a velocity of
+75~\kms\ (FWHM=8.5~\kms). This is illustrated in Fig.~\Fspecial b/d. The dashed
lines show the \H2, \HI\ and \OI\l1039.230 line at $-$20~\kms, while the dotted
lines indicate the +75~\kms\ component.
\par No \HI\ emission is seen at +75~\kms\ in the Leiden-Dwingeloo Survey (LDS)
of Hartmann \& Burton (1997) (36\arcmin\ beam). However, an Effelsberg 21-cm
spectrum (9\arcmin\ beam) toward this target shows a weak
($N$(\HI)=6.4\tdex{18}\,\cmm2), narrow (FWHM=8.2~\kms) component at a velocity
of +73~\kms\ (Fig.~\Fspecial d). Note that FWHM(\HI)$\sim$FWHM(\H2). This may
represent a very small cloud or even a circumstellar shell, so that beam
dilution reduces the 21-cm brightness temperature to be below the detection
limit of the LDS. \HI\ with similar velocities is not detected in any other
direction near Mrk\,153 ($l$=156\fdg73, $b$=56\fdg01). On the other hand, in the
HVC catalog of Wakker \& van Woerden (1991) small (compact) \HI\ clouds with
$v_{\rm LSR}$$\sim$80--110~\kms\ are common at $l$$>$170\deg, $b$$<$60\deg. It
is possible that the object seen toward Mrk\,153 is an outlier in this
population.
\par The ratio of the beam areas of the Dwingeloo and Effelsberg telescopes is
16, so a cloud with a diameter of 9\arcmin\ or less that has $T_B$=0.4~K at
9\arcmin\ resolution would have $T_B$$<$0.025~K in a 36\arcmin\ beam, which is
well below the rms noise of the LDS data (0.05~K). Such a cloud might be
detectable in an 1\arcmin\ resolution interferometer map, however. If the cloud
is smaller than 9\arcmin\ the value of $N$(\HI) derived from the Effelsberg
21-cm brightness temperature scales with the assumed cloud diameter $\alpha$ (in
arcmin) as $N$(\HI)=6.4\tdex{18}\,(81/$\alpha^2$)\,\cmm2.
\par Using an assumed size $\alpha$, and defining $D_{\rm kpc}$ as the cloud's
distance in kpc (so that its diameter $d$=$\alpha D_{\rm kpc}$), the volume
density in this cloud is $n_{\rm H}$=$N$(\HI)/$d$ =
577\,$\alpha^{-3}$\,$D_{\rm kpc}^{-1}$\,\cmm3. In formation-dissociation
equilibrium, the \H2\ and \HI\ volume densities are related by:
\begin{equation} % 1
{ n({\rm H}_2) \over n({\rm H\,I}) }\ =\ { <k>\,\beta_0 \over n_{\rm H}\,R }
\end{equation}
(Spitzer 1978), where $<$$k$$>$=0.11 is the probability that the \H2\ molecule
is dissociated after photoabsorption, $\beta_0$ is the photoabsorption rate per
second, and $R$ is the \H2\ formation rate in cm$^3$\,s$^{-1}$. Assuming that
the formation rate in halo gas is similar to that in disk gas, defining $\chi$
as the strength of the interstellar UV radiation field relative to that near the
Sun, and defining $\phi$ as the fraction of the \HI\ gas in the line of sight
through the cloud that is physically related to the molecular hydrogen, Richter
et al.\ (2003) turn this equation into:
\begin{equation} % 2
n_{\rm H}\ \sim\ 1.8\times10^6\ {N({\rm H}_2)\over N({\rm H\,I})}\ {\chi\over \phi}\ {\rm cm}^{-3}.
\end{equation}
According to the model of the Galactic radiation field presented by Fox et al.\
(2005), at a latitude of $b$=56\deg, and for large distances from the Sun, the
ionizing radiation field drops as about $\chi$$\sim$0.03/$D_{\rm kpc}$. Then,
$n_{\rm H}$=1.4\,$\alpha^2$\,$D_{\rm kpc}^{-1}$\,$\phi^{-1}$~\cmm3. Combining
all these results yield $\alpha^5$=412$\phi$. Since $\phi$$<$1, this means that
formation-dissociation equilibrium implies $\alpha$$<$3.3 arcmin. This in turn
would mean $N$(\HI)$>$4.8\tdex{19}\,\cmm2, $n_{\rm H}$$<$15\,$D_{\rm
kpc}^{-1}$~\cmm3, and $d$=1.1\,$D_{\rm kpc}$\,pc.
\par The +75~\kms\ cloud is also seen in three metal lines. \OI\l1039.230 has an
equivalent width of 65$\pm$10\,m\AA, EW(\NI\l1134.980)=23$\pm$9\,m\AA, and
EW(\SiII\l1020.744)=43$\pm$11\,m\AA. Assuming an intrinsic FWHM of 8.2~\kms, and
using the 21-cm \HI\ measurement (the Lyman lines in the FUSE spectrum are too
confused) these equivalent widths imply column densities
$N$(\OI)= 8.2$^{+18}_{-3}$\tdex{15}\,\cmm2,
$N$(\NI)= 6.3$^{+4.2}_{-3.0}$\tdex{13}\,\cmm2,
$N$(\SiII)= 7.2$^{+8.4}_{-3.7}$\tdex{14}\,\cmm2.
which correspond to abundances
$A$(\OI)= $-$0.41\,+\,2\,$\log(\alpha/3.3)$,
$A$(\NI)= $-$1.80\,+\,2\,$\log(\alpha/3.3)$,
$A$(\SiII)= $-$0.35\,+\,2\,$\log(\alpha/3.3)$.
Since $\alpha$ should be $<$3.3 arcmin, this implies a subsolar abundance for
this cloud. Although there is no strong evidence, a strongly subsolar ($<$$-$1)
abundance seems unlikely, so $\alpha$ is probably $>$1 arcmin. If the \H2\ and
\HI\ are well-mixed, this cloud may be a low density (100--1\,\cmm3), 0.1--10~pc
diameter cloud in the Galactic Halo ($D$=0.1--10~kpc), with slightly subsolar
abundances.

\medskip
\par{\it Mrk\,876.} This sightline passes through several especially interesting
\HI\ structures, including two HVC complex~C components at $-$173 and
$-$133~\kms\ and the Draco Nebula at $-$30~\kms. The metal-line absorption
associated with the HVC was studied by Murphy et al.\ (2000), who used the
CalFUSE v1.8.7 calibration and only the first observation (P1073101). The head
of the Draco Nebula was studied in detail in several papers by Mebold and
co-workers (see Mebold et al.\ 1985, Herbstmeier et al.\ 1993 and references
therein), who argued for a distance of about 1~kpc ($z$$\sim$0.5~kpc) for this
cloud. There is no \H2\ absorption at the velocity of the HVC, down to a limit
of log\,$N$(\H2)=14.06, but the Draco Nebula is clearly detected in \H2,
although it blends with the low-velocity ($v$=$-$2~\kms) component. Example \H2\
lines are shown in Fig.~\Fspecial c. This is also a case where without very
careful alignment of the different segments the intermediate- and low-velocity
\H2\ would blend too much to measure separately, especially since the first
exposure of Mrk\,876 is one of the few with a remaining slope in the wavelength
scale, and not accounting for this would smear out the IVC \H2\ at the edges of
the LiF1A and LiF2B segments.
\par The implied FWHM of the \H2\ absorption is 8.8~\kms, smaller than the
14.7~\kms\ seen in the 9\arcmin\ resolution Effelsberg \HI\ spectrum. The \HI\
column density in the Effelsberg data is 6.9\tdex{19}\,\cmm2. Combining this
with log\,$N$(\H2)=15.47, Eq.~4 implies a density of 77\,$\chi/\phi$ if the \H2\
and \HI\ are in equilibrium. Although the Draco Nebula appears to be relatively
close to the plane ($z$$\sim$500~pc), it is above much of the dust and thus
$\chi$ probably lies somewhere between 0.05 and 0.5. Since $\phi$ is probably
$\sim$1 in this case, a hydrogen density of 4--40\,\cmm3\ is implied.

\medskip
\par{\it NGC\,1068, NGC\,3310.} These sightlines are special cases in that they
need to be smeared with a gaussian having an FWHM of 45~\kms\ (NGC\,1068) or
80~\kms\ (NGC\,3310). This is easily justified, however, because both objects
have extended ($\sim$30\arcsec) UV emission that basically fills the \FUSE\ LWRS
aperture. Toward NGC\,1068 the $J$=2 and $J$=3 column densities can still be
determined using a curve-of-growth, but a 100~\kms\ wide window is needed. In
both sightlines the $J$=0 and $J$=1 column densities require Voigt profile fits.
Toward NGC\,3310 the fitting is rather difficult, as there are two \HI\ clouds
in the line of sight -- the IV-Arch at $-$47~\kms\
($N$(\HI)=6.4\tdex{19}\,\cmm2), and local gas at $-$17~\kms\ with
$N$(\HI)=7.7\tdex{19}\,\cmm2\ (see Wakker et al.\ 2001). Both of these appear to
have relatively high $N$(\H2): 7.8\tdex{18} and 5.4\tdex{18}\,\cmm2,
respectively. Thus, a special method was used to determine $N(J)$. First, a
combined curve-of-growth fit was made to the $J$=2 and $J$=3 lines, separately
for each component. This yields a $b$-value. Next the unblended positive- and
negative-velocity wings of $J$=0 and $J$=1 lines were fitted to estimate $N$(0)
and $N$(1) in both components. Then \T01 was set for the implied $N$(\H2), using
the $N$(\H2) vs \T01 correlations discussed in \Sect\Scorrel. Finally, \T23 and
\T34 were assumed to be equal to \T01 -- without this assumption the theoretical
profiles for $J$=2 and $J$=3 lines are too deep. Figure~\Fspecial e shows some
examples of \H2\ absorption lines toward NGC\,3310. The thick lines show the
profiles using an FWHM=80~\kms\ gaussian. The thin lines show the result that
would be obtained with the usual 20~\kms\ resolution.
\par The high \H2\ column densities toward NGC\,3310 are unusual, but they are
implied by the data. A map of the IVC using the LDS data shows that the
sightline passes through a rather small ($<$1\deg) core. It is possible that the
extended UV source picks out a tiny piece of the IVC (and also of the
low-velocity gas) that does have high $N$(\H2). This could be completely
accidental, or it could mean that small high-density molecular clouds are
common, and will be seen against extended UV sources. A possible way to confirm
or deny this would be to obtain several spectra of NGC\,3310 with the \FUSE\
MDRS or HIRS slit placed at different positions.

\medskip
%      244.000  5.28633e+18     0.283634     0.359488      7.70000     110
%     113     252      1.00000      350.000      1.40000     -30.0000
% for J=0,1 => 244.000  6.91e+16 0 0 20.0 133 241
% for J=2,3 => 244.000  2.91e+16 0 0 20.0 133 215
\par{\it NGC\,3783.} The direction toward NGC\,3783 contains a high-velocity
cloud at about +240~\kms, which has been relatively well studied (West et al.\
1985; Lu et al.\ 1994, 1998; Sembach et al.\ 2001a; Wakker et al.\ 2002). This
HVC (HVC\,287+23+240 or WW\,187) has metal abundances $\sim$0.25 times solar and
is interpreted as a core in the Leading Arm of the Magellanic Stream. Sembach et
al.\ (2001a) reported strong \H2\ absorption in this cloud, with 30
uncontaminated lines seen. Their analysis produced the following parameters:
log\,$N$(0,1,2,3,4)= 16.24$\pm$0.20, 16.64$\pm$0.10, 15.20$\pm$0.08,
14.80$\pm$0.08, $<$14.6, which corresponds to log\,$N$(\H2)=16.80$\pm$0.10,
\T01=133$^{+37}_{-21}$~K, \T23=241$^{20}_{-17}$~K and FWHM=20~\kms. The analysis
presented in this paper yields log\,$N$(0,1,2,3,4)= 17.67$\pm$0.70,
18.00$\pm$0.70, 16.60$\pm$0.50, 15.49$\pm$0.20, 14.00$\pm$0.13,
log\,$N$(\H2)=18.17, FWHM=8.2~\kms, \T01=117~K, \T23=127~K.
\par The main reason for the difference in these column densities is the very
different conclusion reached concerning the intrinsic FWHM (20 vs 8.2~\kms).
This seems to be caused by the fact that Sembach et al.\ (2001a) decided that
the weakest $J$=2 and $J$=3 lines (L(2-0) P(2) $\lambda$1081.267, L(1-0) P(2)
$\lambda$1096.438, L(1-0) R(3) $\lambda$1096.725, L(1-0) P(3) $\lambda$1099.787)
were not detected, as can be seen from their table of equivalent widths. In
Fig.~\Ftset\ some examples of the profiles implied by the two descriptions are
compared. Those implied by the \H2\ column densities and FWHM in Table~\Tmolhyd\
shown as thick solid lines, while the ones implied by the values given by
Sembach et al.\ (2001a) are shown as thin solid lines. Clearly, for $J$=0, $J$=1
and for strong $J$=2 and $J$=3 lines, the two prescriptions produce almost the
same result. There are some subtle differences in detail, such as near
\vlsr=190~\kms\ in the W(0-0) Q(1) $\lambda$1009.771 and L(4-0) P(1)
$\lambda$1051.033 panels, both showing that the thick lines give a slightly
better prediction. The main difference is that the low-$b$, high-$N$ profile is
clearly much better for the L(1-0) R(2) $\lambda$1094.244, L(1-0) P(2)
$\lambda$1096.438, L(1-0) R(3) $\lambda$1096.725, and L(1-0) P(3)
$\lambda$1099.787 lines. The second and third of these four mix with the low-,
intermediate- and high-velocity \FeII\ lines, but by comparing the profile with
other \FeII\ lines it is clear that the high-velocity \H2\ lines are not blended
with \FeII.
\par It should be noted that the Sembach et al.\ (2001a) analysis was based on
just the first 37~ks exposure of this object, whereas the current analysis uses
all seven exposures (totaling 196~ks). However, even in the first exposure the
weakest $J$=2 and $J$=3 absorption lines clearly show the high-velocity
component.
\par This comparison provides a prime illustration of how difficult it is to
determine \H2\ column densities in the range log\,$N$(\H2)=16.5 to 18.2. For
strong lines there is little difference in a low-$b$, high-$N$ vs a high-$b$,
low-$N$ profile, since the lines have not yet developed damping wings. Only the
weakest $J$=2/3 lines contain some information that allows one to constrain $b$,
and the final result is rather sensitive to those weak lines.
\par Wakker et al.\ (2002) found that the \HI\ column density in the HVC toward
NGC\,3783 is 8.3\tdex{19}\,\cmm2, implying a molecular fraction
$f$(\H2)=2\,$N$(\H2)/(2\,$N$(\H2)+$N$(\HI)) of 0.03. Since this cloud is likely
to be part of the Leading Arm of the Magellanic Stream, at a distance of 50 to
100~kpc, the hydrogen volume density implied by Eq.~4 is
$\sim$10\,($D$/100~kpc)\,$\phi^{-1}$. This can be compared to the volume
densities of 20~\cmm3\ implied by the high-resolution (1\arcmin) \HI\ data. The
high-resolution \HI\ maps of Wakker et al.\ (2002) show that NGC\,3783 does not
lie in the direction of a bright subcore in the HVC, so the fact that there
still is a substantial amount of \H2\ suggests that $\phi$ is large. Thus, it
appears likely that the \H2\ and \HI\ in this cloud are in
formation-dissociation equilibrium, contrary to the conclusion of Sembach et
al.\ (2001a). 

%%%%%%%%%%%%%%%%%%%%%%%%%%%%%%%%%%%%%%%%%%%%%%%%%%%%%%%%%%%%%%%%%%%%%%%%%%%%%%%%

\subsection{Molecular fraction}
\par Figure~\FHIHt\ a presents a plot of the average molecular fraction in each
sightline, defined as
\begin{equation} % 1
f({\rm H}_2)\ =\ {2\,N({\rm H}_2) \over 2\,N({\rm H}_2) + N({\rm H}\,I) }.
\end{equation}
Solid points are for low-velocity gas, i.e.\ only \HI\ and \H2\ components at
absolute velocities less than 30~\kms\ were used. Open points are for the
intermediate-velocity clouds (see also next subsection). 
\par It should be stressed strongly that for the low-velocity gas these $f$(\H2)
values have no direct physical meaning. It is almost guaranteed that each
sightline contains several gas clouds with similar velocity, containing both
\HI\ and \H2. At the available spectral resolution (and also considering the
intrinsic linewidths), the \H2\ absorptions of these clouds cannot be separated.
Only the total can be measured. Each of this unknown number of clouds
contributes an unknown fraction of the \HI\ and \H2\ to the total column density
that is observed. Any attempt to compare $N$(\HI) and $N$(\H2) directly and
derive the local fraction of nuclei that is in the form of \H2\ will run afoul
of this fact. Thus, the usual study of how the molecular fraction varies with
\HI\ column density (e.g.\ Savage et al.\ 1977) can not be done for the
sightlines in the \FUSE\ extragalactic sample. Such a study is justified for
sightlines to nearby disk stars, where the sightline intersects just one ISM
cloud. What can be done is to study how important the \H2\ is as a major
constituent of the local ISM.
\par The $f$(\H2) vs $N$(\HI) plot shows three regions. a) The four sightlines
with log\,$N$(H)$>$20.75 are ESO\,141$-$G55, HS\,0624+6907, Mrk\,1095 and
NGC\,3783, and they represent four of the nine sightlines at latitudes below
27\deg. It is interesting that no small values of $f$(\H2) are found at high
hydrogen column densities, suggesting that at least one of the clouds in the
sightline is dense enough have a high molecular fraction.
\par b) At log\,$N$(H)$<$20.2 the average molecular fraction mostly lies between
\dex{-3.3} and \dex{-5.8}. The majority of these points correspond to
intermediate-velocity clouds (open circles). In this case the comparison {\it
is} physically meaningful, and it implies that log\,$N$(H) has to be larger than
$\sim$20.2 before a transition to \H2\ takes place. The two unusual points at
(19.9,$-$0.9) are for NGC\,3310 (where $N$(\H2) is very difficult to determine),
and for PG\,1351+640.
\par c) In the intermediate region (log\,$N$(H)=20.2 to 20.7) a large range is
seen for $f$(\H2). It is likely that for sightlines with low $f$(\H2) the \HI\
column density is spread over several clouds, while for sightlines with high
$f$(\H2) most of the \HI\ column density is concentrated in one high column
component.

%%%%%%%%%%%%%%%%%%%%%%%%%%%%%%%%%%%%%%%%%%%%%%%%%%%%%%%%%%%%%%%%%%%%%%%%%%%%%%%%

\subsection{IVCs}
\par Table~\TIVC\ lists the \H2\ parameters found for intermediate-velocity gas,
both in this paper and in Richter et al.\ (2003). This shows that similar
answers are obtained for 6 of the 18 sightlines (3C\,273.0, Mrk\,59, Mrk\,509,
Mrk\,876, PG\,1259+593, PG1351+640). In four sightlines (Mrk\,279, Mrk\,421,
Mrk\,1513, NGC\,7714) the IVC only popped up in the additional data that was
obtained after May 2002, i.e.\ data not included by Richter et al.\ (2003). In
two sightlines (Mrk\,817 and NGC\,3783) the analysis presented here revealed an
IVC component that was not found by Richter et al.\ (2003). The $-$45~\kms\
component toward NGC\,3783 is not seen in \HI, although it appears needed to fit
some of the stronger \H2\ lines. Still, the reality of this component remains in
some doubt. New IVC components are found toward three sightlines not included in
the Richter et al.\ (2003) sample: Mrk\,153 (see above), NGC\,4051 and
1H\,0717+714.
\par Results that differ substantially from those listed by Richter et al.\
(2003) are found for three sightlines -- NGC\,3310, NGC\,4151 and PG\,1116+215.
The difficulties with NGC\,3310 were described above. Toward NGC\,4151 the \HI\
spectrum is complex, with narrow peaks at $-$41 and $-$21~\kms\ on top of a
broad base (see Wakker et al.\ 2003). A very good \H2\ fit can be obtained with
a single component at $-$21~\kms. Richter et al.\ (2003) identified this \H2\ as
associated with the broad IV-Arch component at $-$29~\kms. Finally, the
PG\,1116+215 sightline is noteworthy for the fact that it is the only sightline
toward which the intermediate-velocity \H2\ is much stronger than the
low-velocity \H2. The FWHM found by Richter et al.\ (2003) is twice as large as
the value found here (13.2 vs 6.5~\kms). This results in column densities of
15.27 vs 16.01, respectively. The higher-$b$, lower-$N$ combination yields
similar theoretical profiles for strong lines, but clearly underpredicts the 3
to 5 weakest lines of each J level, as illustrated by Figs.~\Fspecial g and h.
This is not the case for the lower-$b$, higher-$N$ combination, which must
therefore be preferred.
\par Molecular hydrogen is not seen in all IVCs. Richter et al.\ (2003) studied
a total of 61 IVC components in 56 sightlines, and report 14 clear and 17
tentative detections. Their sample included many sightlines with much lower S/N
ratio than the limit in this paper. Figure~\FHIHt b presents a scatter plot of
log\,$N$(\HI) vs log\,$N$(\H2) in the \ivlos\ IVCs in the current sample. \H2\
is not detected for $N$(\HI)$<$19.2, but is seen in 8 of 20 cases (40\%) with
$N$(\HI) between 19.25 and 19.75, and in 6 of 9 (66\%) with $N$(\HI)$>$19.75.
Thus, the fraction of sightlines with intermediate-velocity \H2\ appears to
increase with $N$(\HI).
\par The IVCs in which \H2\ is detected include six sightlines through the IV
Arch, two through the LLIV Arch, as well as five smaller IVCs. As previously
argued by Richter et al.\ (2003), the fact that \H2\ is seen in many IVCs
strongly suggests the presence of dust in these clouds.

%%%%%%%%%%%%%%%%%%%%%%%%%%%%%%%%%%%%%%%%%%%%%%%%%%%%%%%%%%%%%%%%%%%%%%%%%%%%%%%%

\subsection{HVCs}
Richter et al.\ (2001) studied high-velocity \H2\ in two sightlines --
Fairall\,9 through the Magellanic Stream and PG\,1259+593 through complex~C.
Although these two sightlines have comparable amounts of \HI, \H2\ was seen
toward Fairall\,9 log\,$N$(\H2)=16.4, but not toward PG\,1259+593
(log\,$N$(\H2)$<$13.96). This led Richter et al.\ (2001) to conclude that there
is dust in the Magellanic Stream but not in complex~C, which is supported by the
abundance ratios of depleted to undepleted elements in these two objects. That
conclusion was further supported by the detection of \H2\ toward NGC\,3783
reported by Sembach et al.\ (2001a).
\par The high S/N sample of this paper does not include Fairall\,9, but there
are 20 sightlines where high-velocity \HI\ is seen. The only high-velocity \H2\
seen remains that toward NGC\,3783. Nine of the remaining 19 sightlines pass
through complex~C, two through complex~A, two through the Outer Arm, three
through positive-velocity HVCs, two through the trailing part of the Magellanic
Stream and one through a compact HVC. Figure~\FHIHt c shows the upper limits
versus the \HI\ column density, with open squares for the complex~C sightlines,
open triangles for the other HVCs. Although seven of the nine HVCs have \HI\
column densities $>$19.25, i.e.\ comparable to those seen in the IVCs, no \H2\
is seen in complex~C down to limits of log\,$N$(\H2)$<$13.84. The \HI\ column
densities toward the other HVCs are generally lower. Six of the ten have
log\,$N$(\HI)$<$19.25, and by analogy with the IVCs \H2\ might not be expected.
However, none of the four with higher $N$(\HI) show \H2\ either. The implication
is that HVCs seem to have little to no dust. 

%%%%%%%%%%%%%%%%%%%%%%%%%%%%%%%%%%%%%%%%%%%%%%%%%%%%%%%%%%%%%%%%%%%%%%%%%%%%%%%%

\subsection{\H2\ in external galaxies}
There are five detections of \H2\ in external galaxies. These are summarized in
Table~\TIVC\ and described below. Figure~\Fexternal\ shows the wavelength region
between rest wavelengths of 1047.7 and 1060.5\,\AA, but displayed such that the
redshifted \H2\ lines of the five targets are aligned. The dotted lines and
labels indicate the positions of the redshifted metal and \H2\ lines, while the
dashed lines without labels show the positions of the Milky Way lines. The
horizontal bar on the lower left side indicates the velocity of the external
galaxy; it shows how far apart each pair of Galactic and extragalactic lines
is.
\par Four of the five \H2\ extragalactic detections are in nearby galaxies, with
the light provided by hot stars in \HII\ regions. In all cases the \H2\ lines
are weak, and they are probably associated with the diffuse ISM in front of the
\HII\ regions. As the light that is seen is the combined light from many OB
stars, both the absorption and the emission spectrum are a combination of many
sightlines, and the \H2\ absorption is just an average over these. Furthermore,
sightlines with low extinction (and thus low $N$(\H2)) are detected
preferentially, so the \H2\ column densities are not expected to be
representative of the dense ISM in those external galaxies. The \H2\ lines also
are likely to have substructure, but the S/N ratios and complicated continua of
the spectra do not allow a detailed study. Nevertheless, in the end it is
possible to proceed as-if the lines originate in a single cloud and derive
column densities and temperatures. These turn out to fit the same patterns as
the \H2\ lines seen in the Milky Way (see \Sect\Scorrel). However, the resulting
column densities should not be thought of as physically meaningful, but as
``pseudo column densities'' that parametrize the absorption spectrum.
\par The five extragalactic \H2\ detections are now described in more detail,
with Mrk\,205 being the only case where the background light is not provided by
the absorbing galaxy.
\medskip
\par{\it Mrk\,205.} It is well known that the sightline to Mrk\,205 intersects
the disk of the nearby galaxy NGC\,4319 (Bowen et al.\ 1991, 1995; Bowen \&
Blades 1993). The sightline passes just 6~kpc from the center of NGC\,4319
($v_{\rm syst}$=1357~\kms). Only a few other such coincidences between a
FUV-bright AGN and a galaxy are known (Ton\,1480 passes 7~kpc from NGC\,4203,
but Ton\,1480 is only half as bright as Mrk\,205 in the FUV; 3C\,232 is even
fainter and passes 13~kpc from NGC\,3067 -- Keeney et al.\ 2005). \H2\
associated with the disk of NGC\,4319 is clearly detected at a velocity of
1281~\kms. Figure~\Fexternal a shows several Galactic $J$=0, 1, 2 and 3 \H2\
lines, as well as these same lines redshifted by 1281~\kms. For the fitting care
was taken to exclude \H2\ lines for which the Galactic component at $-$5~\kms\
and the NGC\,4319 component blend. The parameters for the \H2\ in NGC\,4319 are
log\,$N$(\H2)=15.78$\pm$0.24, \T01=271~K and \T23=353~K. The expected
temperatures for Galactic \H2\ (see Sect.~\Scorrel) at this column density are
\T01=177~K and \T23=281~K. This corresponds to measured values of $N$(2) and
$N$(3) that are about 0.1 and 0.25 dex higher and $N$(0) 0.16 dex lower than
expected from $N$(1). These differences are not large and thus it is safe to
conclude that the properties of the \H2\ in NGC\,4319 are similar to those in
the Milky Way. \medskip
\par{\it NGC\,604.} Bluhm et al.\ (2003) measured the \H2\ for several
sightlines toward \HII\ regions in M\,33. One of these (NGC\,604) falls in the
high S/N sample of this paper. For this sightline, Bluhm et al.\ (2003) used the
A0860101 dataset, which was obtained through the LWRS aperture; they derived
upper limits of log\,$N$(0,1,2,3) $<$13.7, $<$14.3, $<$14.4, $<$14.0. After that
paper was published NGC\,604 was observed again, but this time through the MDRS
aperture (observation B0180201). Although the flux in the new dataset is lower
(10.6 vs 31.2\tdex{-14}\,\fu) as is the S/N ratio (14.1 vs 17.4 at 1063\,\AA),
the \HII\ region is an extended source, so the narrower aperture results in much
sharper lines. As a result, one $J$=0, three $J$=1, one $J$=2 and three $J$=3 
M\,33 \H2\ lines are detected at the 2 to 6$\sigma$ level, resulting in
approximate column densities of log\,$N$(0,1,2,3)= 14.3, 14.8, 14.3, 14.5.
Figure~\Fexternal b shows several of these weak lines. To reconcile the column
densities implied by the equivalent widths with the shape of the profiles, it is
necessary to assume that the instrumental resolution is 30~\kms. This is most
likely a result of the fact that NGC\,604 is an extended source, which smears
out the image on the detector. Also, \T23 is unusually high considering the
value of $N$(\H2) (see \Sect\Scorrel), which may be related to the fact that the
\H2\ absorption is an average between many sightlines.
\medskip
\par{\it NGC\,625.} The data for NGC\,625 ($v_{\rm gal}$=396~\kms) have a
comparatively low S/N ratio (12.1), and the \H2\ that may seen at a velocity of
396~\kms\ is weak, showing 3 to 8$\sigma$ detections in several $J$=0, $J$=1 and
$J$=3 lines (see Fig.~\Fexternal c for a few examples). It is clear that these
detections are real, even though many lines are a bit ragged around the edges --
that is, they seem to show a little broader absorption. There is no clear
pattern to this, however, in that lines of the same $J$ level show different
substructure. Some of this is probably due to the low S/N ratio, some to stellar
lines in some of the stars that provide the light.
\medskip
\par{\it NGC\,5236 (M\,83) and NGC\,5253.} The detections of extragalactic \H2\
at $v$=500~\kms\ in the spectrum of NGC\,5236 (M\,83) and $v$=404~\kms\ in
NGC\,5253 were first reported by Hoopes et al.\ (2004). The continua of these
objects have many undulations, caused by stellar lines, so that the piecewise
continuum fit requires many small pieces, leaving some \H2\ lines too confused.
This also means that the equivalent widths have large systematic errors. These
are not taken into account in the fitting process, but it means that the real
errors on $N(J)$ are much larger than the quoted values. Figures~\Fexternal d
and e show examples of detected lines. Panel d clearly shows the undulating
continuum toward NGC\,5236, and shows some artifacts associated with the
piecewise fit (e.g.\ near 1055.5\,\AA).
\par For NGC\,5236 Hoopes et al.\ (2004) report only one 1.7$\sigma$
(25.6$\pm$15~m\AA) detection in the L(9-0) R(1) $\lambda$993.018 line. However,
a detailed assessment, and using the carefully aligned and combined LiF1A+LiF2B
spectra shows that five $J$=0, nine $J$=1, four $J$=2 and five $J$=3 lines can
be seen, with typical equivalent widths of 30~m\AA\ ($J$=0, 2, 3) or 50~m\AA\
($J$=1) and typical errors of $\sim$6\,m\AA. The FWHM appears to be 12~\kms, and
the implied column density is $N$(\H2)=1.9\tdex{15}\,\cmm2. Compare this to the
values of 25~\kms\ and 1.8\tdex{15}\,\cmm2\ listed by Hoopes et al.\ (2004).
\par Hoopes et al.\ (2004) report a few $J$=1 and $J$=2 detections in the
spectrum of NGC\,5253. In the combined LiF1A+LiF2B spectra used here, two
unblended lines of $J$=0, four of $J$=1, 2, and nine of $J$=3 are clearly seen,
as are many more blended lines. Detections range from 20 to 70~m\AA, with
typical errors of 7~m\AA. The implied column density is 1.7\tdex{15}\,\cmm2,
with an FWHM of 10~\kms. Hoopes et al.\ (2004) listed 2.2\tdex{15}\,\cmm2.
\par In both galaxies, the theoretical profiles constructed using the column
densities and the FWHM implied by the curve-of-growth need to be smoothed with a
45~\kms\ (NGC\,5236) or 30~\kms\ (NGC\,5253) wide instrumental profile to make
them look similar to the observed profiles. This can be explained by the fact
that the background galaxies are extended sources. For the same reason, the
derived temperatures must represent a mixture of gas in the many sightlines to
individual stars. Yet, the average seems to follow the usual $N$(\H2) vs \T01
and \T23 correlations described in \Sect\Scorrel.
\par Toward NGC\,5236 the Milky Way molecular hydrogen requires special
treatment. It has low column density (log\,$N$(\H2)$\sim$14.8). No $J$=0 or
$J$=2 lines are detected, but six $J$=1 lines and four $J$=3 lines show up.
Assuming the expected \T01 and \T23 for log\,$N$(\H2)=14.8 (see \Sect\Scorrel)
shows that the column densities for these two levels are normal, and that the
$J$=0 and $J$=2 should indeed be undetectable at the target's S/N level. Thus,
the temperatures listed in Table~\Tmolhyd\ are the ones expected from the
correlations.

%%%%%%%%%%%%%%%%%%%%%%%%%%%%%%%%%%%%%%%%%%%%%%%%%%%%%%%%%%%%%%%%%%%%%%%%%%%%%%%%
%%%%%%%%%%%%%%%%%%%%%%%%%%%%%%%%%%%%%%%%%%%%%%%%%%%%%%%%%%%%%%%%%%%%%%%%%%%%%%%%

\subsection{Correlations}
\par Figure~\Fcorrel\ shows scatter plots of many combinations of the measured
\H2\ parameters. Filled circles are for the most reliable measurements, using
sightlines with S/N ratio $>$13 near 1031\,\AA\ (or 1063\,\AA). Fits to
components in sightlines with lower S/N ratio or fits that seemed less reliable
for one reason or another are shown as open circles. Clear correlations exist
between the \H2\ parameters. Least-squares fits were made to three of the
scatter plots, using only the filled circles, giving:
\begin{eqnarray}
T_{01}                =&  -23.7\ (\log N({\rm H}_2) - 14) + 219    \\
\log {N(2)\over N(0)} =& -0.323\ (\log N({\rm H}_2) - 14) + 0.274  \\
\log {N(3)\over N(1)} =& -0.513\ (\log N({\rm H}_2) - 14) - 0.046  \\
\log {N(4)\over N(3)} =& -0.210\ (\log N({\rm H}_2) - 14) - 0.669. \\
\end{eqnarray}
These four fits are shown by the solid black lines in panels d, j, l and o of
Fig.~\Fcorrel. Using these relations, it is possible to predict values for
$N(J)$ given a value for $N$(\H2). The implied relations for the other panels
are shown by the dashed lines. These show that unlike the \T01 correlation, the
correlations between $N$(\H2) and \T23 or \T34 are not linear.
\par In the dense ISM the quantity \T01 is likely to be similar to the kinetic
temperature of the gas, as proton-exchange collisions are the dominant
excitation mechanism. For sightlines with $N$(\H2)$>$\dex{18}\,\cmm2,
$<$\T01$>$=103$\pm$19~K. This can be compared to the average for the nearby
Galactic Disk found by Savage et al.\ (1977) ($<$\T01$>$=77$\pm$17~K), and the
average for the LMC disk found by Tumlinson et al.\ (2002) (82$\pm$21~K).
Clearly, even for the denser high-latitude ISM, $<$\T01$>$ is higher than in the
Disk. A possible explanation is that the average density of gas sampled by the
high-latitude sightlines is lower, and the excitation of the $J$=1 level is no
longer purely thermal.
\par An important conclusion to take away from the correlation between \T01 and
log\,$N$(\H2) is that it is important to specify the \H2\ column density range
when calculating a value for $<$\T01$>$. The fit predicts a value of \T01=81 at
log\,$N$(\H2)=20, comparable to the high-column density average found by Savage
et al.\ (1977). But at log\,$N$(\H2)=15 the expected \T01 is 195~K,
significantly higher. A hint of this effect was already found by Spitzer \&
Cochran (1973) in their sample of sightlines through nearby dense disk gas.
\par Theoretical models of the population level distribution of \H2\ were
presented by Browning et al.\ (2002). They take into account the formation of
\H2\ on dust grains, and the manner in which the higher rotational levels of
\H2\ are excited by UV radiation. Results for several models are presented,
using a standard Galactic radiation field ($\chi$=1), a much stronger field
($\chi$=50), a standard formation rate, and 1/10th the standard formation rate.
These results are only expressed in the form of plots of $N$(\H2) vs
log\,$N$(4)/$N$(2). The few model points that can be obtained from the Browning
et al.\ (2002) paper are overplotted as crosses on panel n of Fig.~\Fcorrel.
This shows that the models a) predict values for $N$(4)/$N$(2) that are in the
correct range at low $N$(\H2) and b) predict an increase in this ratio as
function of $N$(\H2). However, the models tend to overpredict $N$(4) by a factor
$\sim$10 to 100 at the highest \H2\ column densities, and by a factor $\sim$50
at log\,$N$(\H2)=18.

%%%%%%%%%%%%%%%%%%%%%%%%%%%%%%%%%%%%%%%%%%%%%%%%%%%%%%%%%%%%%%%%%%%%%%%%%%%%%%%%

\subsection{Sky map}
\par The distribution of the \H2\ column densities on the sky is shown in
Figs.~\FskymapN\ and \FskymapS. In this representation, each position on the sky
is given a color that depends on the value of $N$(\H2) toward the nearest
background target, up to 12 degrees away. Figure~\FskymapN\ is for the northern
Galactic sky, Fig.~\FskymapS\ for the southern sky. If an upper limit is found,
the area around the background source is only colored out to a radius of 3\deg.
Only \H2\ components with \vlsra$<$25~\kms\ are included in these figures. In
the northern sky, $N$(\H2) is low ($<$16.6) for most latitudes above about
45\deg, the exception being the sightline toward PG\,1211+143. In the southern
sky the sightlines with higher $N$(\H2) concentrate around $l$=90\deg, and there
does not seem to be a strong gradient with latitude. In all these sightlines
\H2\ is a trace constituent of the ISM, as $N$(\HI) is $>$8\tdex{19}\,\cmm2\
everywhere.

%%%%%%%%%%%%%%%%%%%%%%%%%%%%%%%%%%%%%%%%%%%%%%%%%%%%%%%%%%%%%%%%%%%%%%%%%%%%%%%%
%%%%%%%%%%%%%%%%%%%%%%%%%%%%%%%%%%%%%%%%%%%%%%%%%%%%%%%%%%%%%%%%%%%%%%%%%%%%%%%%

\section{Conclusions}
\par This paper presents measurements of \H2\ column densities toward
\finalnumberofsources\ extragalactic targets observed with \FUSE\ for which the
final combined spectrum has signal-to-noise ratio $>$\snlimit. The individual
observations were calibrated with the CalFUSE calibration pipeline version 2.1
or 2.4. Final shifts were found for each detector segment, in order to align the
metal and \H2\ lines with either a 21-cm \HI\ spectrum or a UV spectrum taken
with the \STIS\ E140M grating on \HST. \H2\ column densities or column density
limits were then measured for the low-, intermediate- and high-velocity \HI\ in
each sightlines, for each rotational level $J$=0 to $J$=4. For weaker \H2\ lines
($N(J)$$\ltsim$16.5) a curve-of-growth was used to estimate $b$ and $N$, while
Voigt profiles were fit at larger column densities.
\par These determinations give the following conclusions:
\par (1) \H2\ is seen in almost all (\finaldetections\ out of
\finalnumberofsources) sightlines toward high-latitude (\babs$>$20\deg)
extragalactic targets. 
\par (2) In the northern galactic hemisphere $N$(\H2) is generally below
\dex{17}\,\cmm2\ at latitudes above 45\deg. In the southern hemisphere $N$(\H2)
shown no correlation with latitude, but instead is higher for $l$=40 to 150\deg.
\par (3) There are strong correlations between $N$(\H2) and the level population
ratios, with the average value of \T01 ranging from 81~K at log\,$N$(\H2)=20 to
219~K at log\,$N$(\H2)=14.
\par (4) As is theoretically expected, the level populations of the higher
rotational states show a relative increase at lower \H2\ column densities, with
\T23 and \T34 rising to 375~K and 575~K at log\,$N$(\H2)=15.
\par (5) An unusual cloud is seen at a velocity of +75~\kms\ toward Mrk\,153.
This cloud has narrow lines (FWHM$\sim$8.5~\kms), and must be small, as it is
detected in a 9\arcmin\ 21-cm beam, but not in a 36\arcmin\ beam. It is detected
in \OI, \NI, and \SiII\ lines. It may be a circumstellar shell, or it may be a
low density (100--1\,\cmm3), 0.1--10~pc diameter cloud in the Galactic Halo
($D$=0.1--10~kpc) with subsolar abundances.
\par (6) The high-velocity \H2\ absorption toward NGC\,3783 that was reported by
Sembach et al.\ (2001a) is shown to have an \H2\ column density of 18.17, rather
than 16.80, because the derived FWHM is 8.2~\kms, rather than the 20~\kms\
reported by Sembach et al.\ (2001a). This factor 25 difference in column
densities only leads to noticeable differences in the \H2\ absorption profiles
for the weakest $J$=2 and $J$=3 lines, which were reported as non-detections by
Sembach et al.\ (2001a). The implication of the higher value for $N$(\H2) is
that the \H2\ in the HVC may be in formation-dissociation equilibrium with \HI\
after all, rather than being remnant \H2\ formed before the tidal interaction.
\par (7) A comparison of the \H2\ in IVCs with the results of Richter et al.\
(2003) shows that in all but three cases similar results are found. Nine new
detections of intermediate-velocity \H2\ are reported here.
\par (8) Intermediate-velocity \H2\ is detected in about half of the cases where
intermediate-velocity \HI\ is seen, with the fraction increasing toward higher
$N$(\HI).
\par (9) Except for one sightline through the Magellanic Stream (NGC\,3783)
high-velocity \H2\ is not detected in the 19 high-velocity \HI\ clouds
intersected by the 73 sightlines, indicating there is little or no dust in HVCs.
\par (10) Molecular hydrogen is seen in the disks of five external galaxies:
NGC\,4319, M\,33, NGC\,625, NGC\,5236 (=M\,83), and NGC\,5253. The detection in
NGC\,4319 is at a distance of 7~kpc from the center of that galaxy, using the
Seyfert Mrk\,205 as a background target, giving log\,$N$(\H2)=15.783, and
relatively normal Boltzmann temperatures. In the other four cases the detection
is the average of many sightlines against OB stars with relatively low
extinction, so the low values of $N$(\H2) are not representative of the ISM in
those galaxies.

\acknowledgements
I thank Marilyn Meade for (re)calibrating the many \FUSE\ datasets. This work
was supported by NASA grants NAG5-9179 (LTSA), NNG04GD85G (ADP), NAG5-13687
(\FUSE) and NNG04GA39G (\FUSE).

%%%%%%%%%%%%%%%%%%%%%%%%%%%%%%%%%%%%%%%%%%%%%%%%%%%%%%%%%%%%%%%%%%%%%%%%%%%%%%%%
%%%%%%%%%%%%%%%%%%%%%%%%%%%%%%%%%%%%%%%%%%%%%%%%%%%%%%%%%%%%%%%%%%%%%%%%%%%%%%%%

%%%%%%%%%%%%%%%%%%%%%%%%%%%%%%%%%%%%%%%%%%%%%%%%%%%%%%%%%%%%%%%%%%%%%%%%%%%%%%%%
%%%%%%%%%%%%%%%%%%%%%%%%%%%%%%%%%%%%%%%%%%%%%%%%%%%%%%%%%%%%%%%%%%%%%%%%%%%%%%%%

\def\captionA{Part of the spectrum for 12 sightlines representative for the
range of observed \H2\ column densities. This 9\,\AA\ range contains lines of
$J$=0, 1, 2, 3, and 4, whose wavelengths are shown by the numerical labels along
the bottom axes. Overplotted on the data are theoretical profiles calculated
using the \H2\ parameters listed in Table~\Tmolhyd. The number in parentheses
after the name of each target gives log $N$(\H2).}

\def\captionB{A series of examples for sightlines that are special cases. In
all panels the low-velocity \H2\ lines are indicated by dashed vertical lines,
and intermediate-velocity \H2\ absorption is shown by dotted vertical lines. (a)
This panel shows that toward 3C273.0 a $-$9~\kms\ \H2\ component is present,
although Sembach et al.\ (2001a) claimed it was absent. (b, d) A narrow \HI\
component is seen at 73~\kms\ toward Mrk\,153. This component also shows up in
\OI\l1039.230 and in \H2. (c) A special, well-studied IVC, the Draco Nebula
($v$=$-$30~\kms), is detected toward Mrk\,876 -- it is seen as a weaker
secondary component on the left of each \H2\ line. (e). The two-component
theoretical profiles toward NGC\,3310 need to be smoothed by 80~\kms\ in order
to represent the data, as shown by the thick lines; smoothing with the usual
20~\kms\ (thin lines) is clearly wrong. (g) This panel illustrates the highest
column density (log\,$N$(\H2)=16.45) intermediate-velocity \H2\ toward
PG\,1351+640). (g, h) These two panels compare the column densities and
$b$-values for PG\,1116+215 listed in Table~3 with those given by Richter et
al.\ (2003); thick lines show the profiles implied by the parameters listed in
Table~\Tmolhyd, while the thin lines show the solution given by Richter et al.\
(2003). See \Sect\SIVC\ for more details.}

\def\captionC{Examples of data and model fits for twelve \H2\ lines in the
spectrum of NGC\,3783. The histograms show the data. Thick lines are the
theoretical profiles implied by the high column density fit to the HVC at
+244~\kms\ that is preferred in this paper (log\,$N$(\H2)=18.17, FWHM=8.2~\kms).
Thin lines show the theoretical profiles implied by the fitted parameters in
Sembach et al.\ (2001) (log\,$N$(\H2)=16.80, FWHM=20~\kms). Clearly, the two
predictions are very similar for most lines, but the Sembach et al.\ (2001)
value underpredicts the absorption in the weakest $J$=2 and $J$=3 lines (panels
h, i, k, l).}

\def\captionD{Panel (a) shows the correlation between total hydrogen column
density and the fraction of \H2. Stars are for low-velocity gas at low latitudes
(\babs$<$30\deg), closed circles for low-velocity gas at higher latitudes,
triangles are for the sightlines with upper limits to $N$(\H2), and open circles
are for intermediate-velocity clouds. Panels (b) and (c) give the correlation
between log\,$N$(\HI) and log\,$N$(\H2) in intermediate-velocity clouds (b) and
high-velocity clouds (c). Closed circles/squares show detections. Open
circles/squares/triangles show upper limits, with squares for complex~C and
triangles for other HVCs.}

\def\captionE{This figure shows lines in the L(4-0) band detected in five
external galaxies. The spectra are aligned on the extragalactic \H2\ line, which
are labeled in each panel, and indicated by the dotted vertical lines. The
dashed vertical lines show the positions of the Milky Way absorption lines. On
the lower left sides, each panel also contains a horizontal bar, which shows the
distance between each pair of Galactic and extragalactic absorption lines.}

\def\captionF{Scatter plots of the measured \H2\ parameters. Filled circles are
for the most reliable measurements (all in sightlines with S/N ratio $>$13),
while open circles are for the remaining ones. The solid lines in panels d, j, l
and o show the least-squares fits made to the filled circles in these plot. The
dashed lines show the relations implied for the other scatter plots. Crosses in
panel n show theoretical predictions from Browning et al.\ (2002). Note that
panels d and i (\T01 or log\,$N$(1)/$N$(0) vs log\,$N$(\H2)) really display the
same correlation, as do panel pairs e/m and f/o. Both are shown for
convenience.}

\def\captionG{Sky map of log\,$N$(\H2) for the Northern Galactic sky. Each
position on the sky is given a a color that depends on the value of $N$(\H2)
toward the nearest background target, up to 12 degrees away (or 3\deg\ for upper
limits). Only components with \vlsra$<$25~\kms\ are included.}

\def\captionH{Sky map of log\,$N$(\H2) for the Southern Galactic sky. See
Fig.~\FskymapN\ for description.}

\newpage\par\begin{figure}\plotfiddle{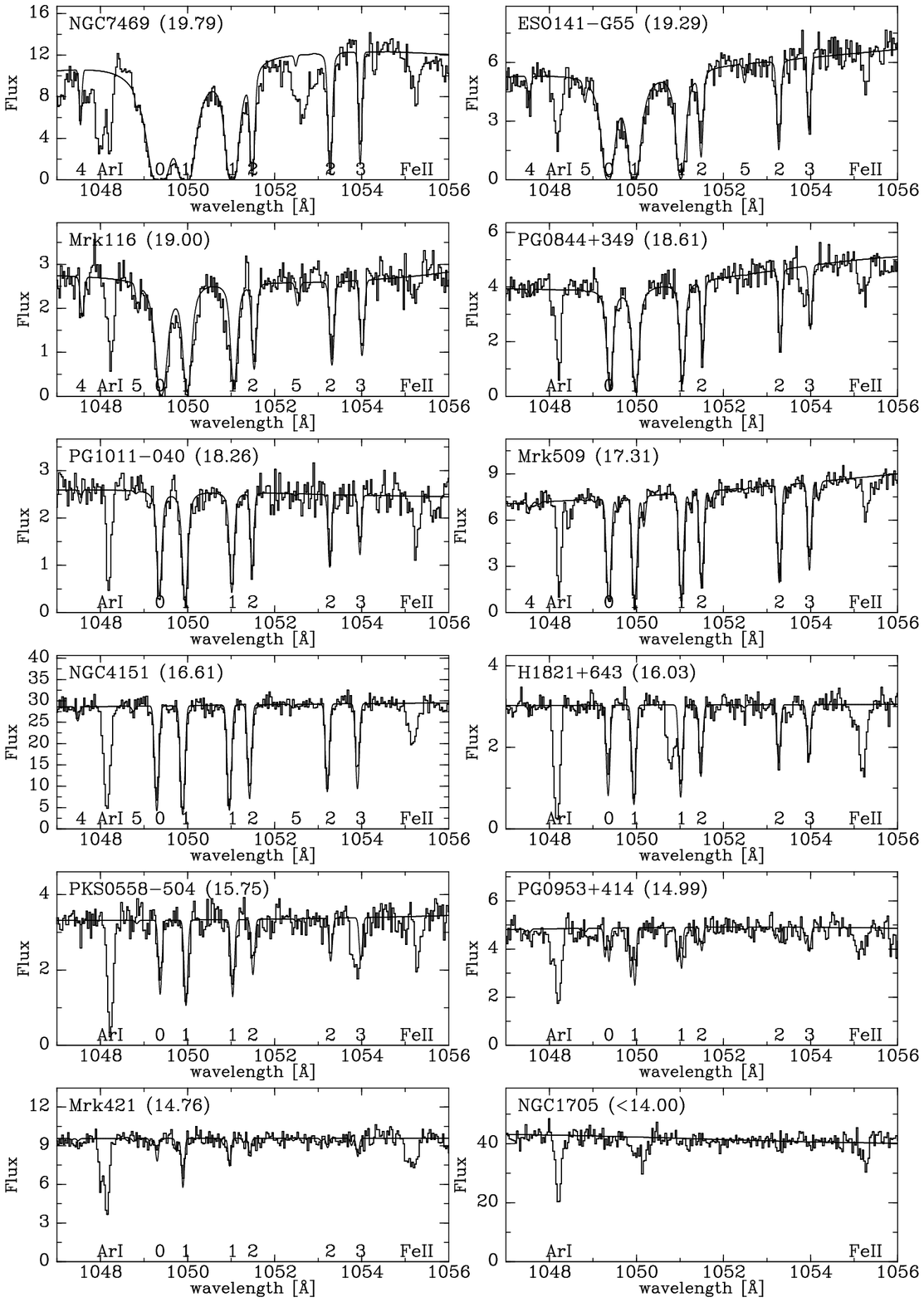}{0in}{0}{345}{490}{0}{00}\figurenum{1}\caption{\captionA}\end{figure}
%newpage\par\begin{figure}\plotone{f1.eps}\figurenum{1}\caption{\captionA}\end{figure}
\newpage\par\begin{figure}\plotone{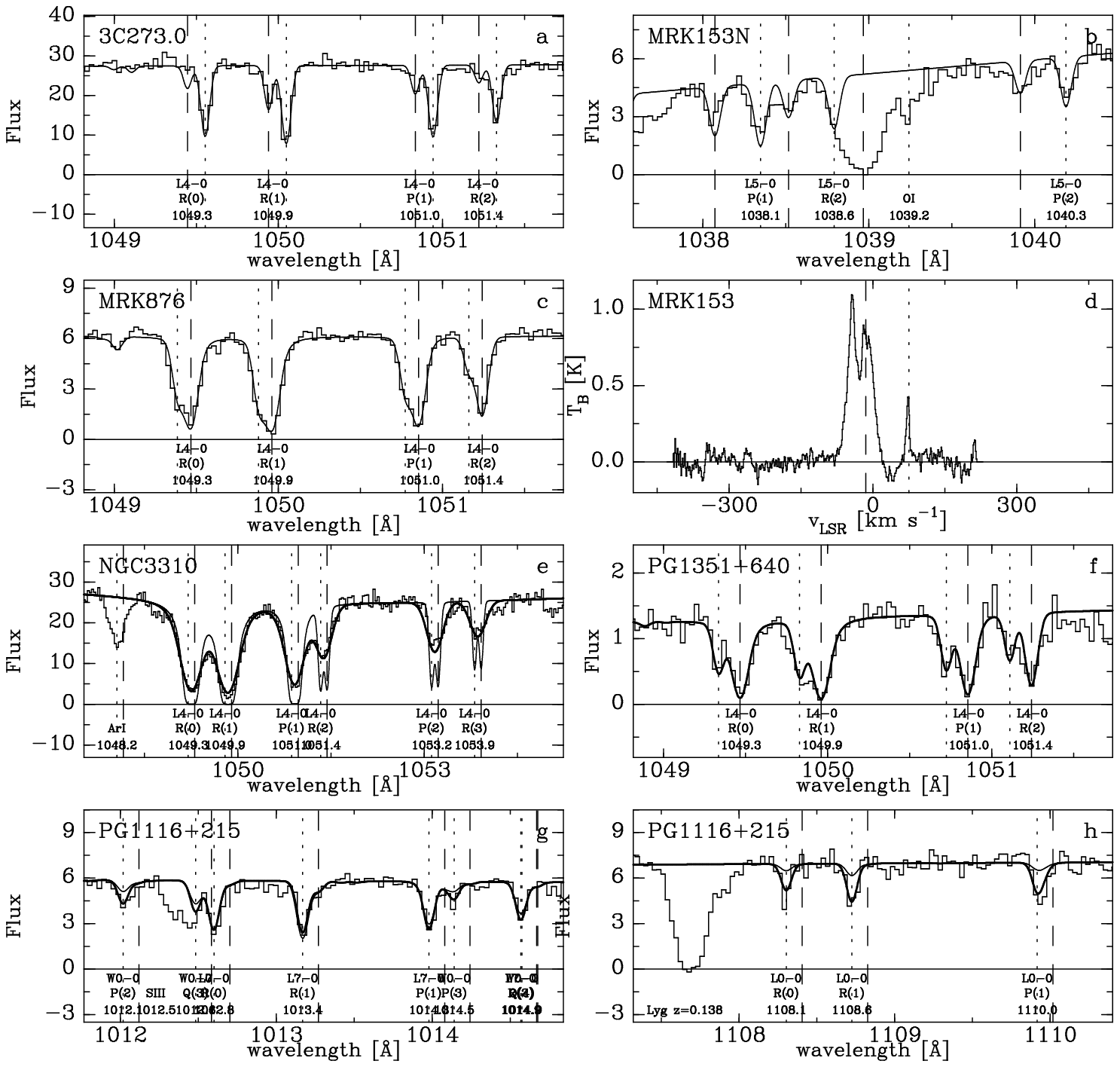}\figurenum{2}\caption{\captionB}\end{figure}
\newpage\par\begin{figure}\plotone{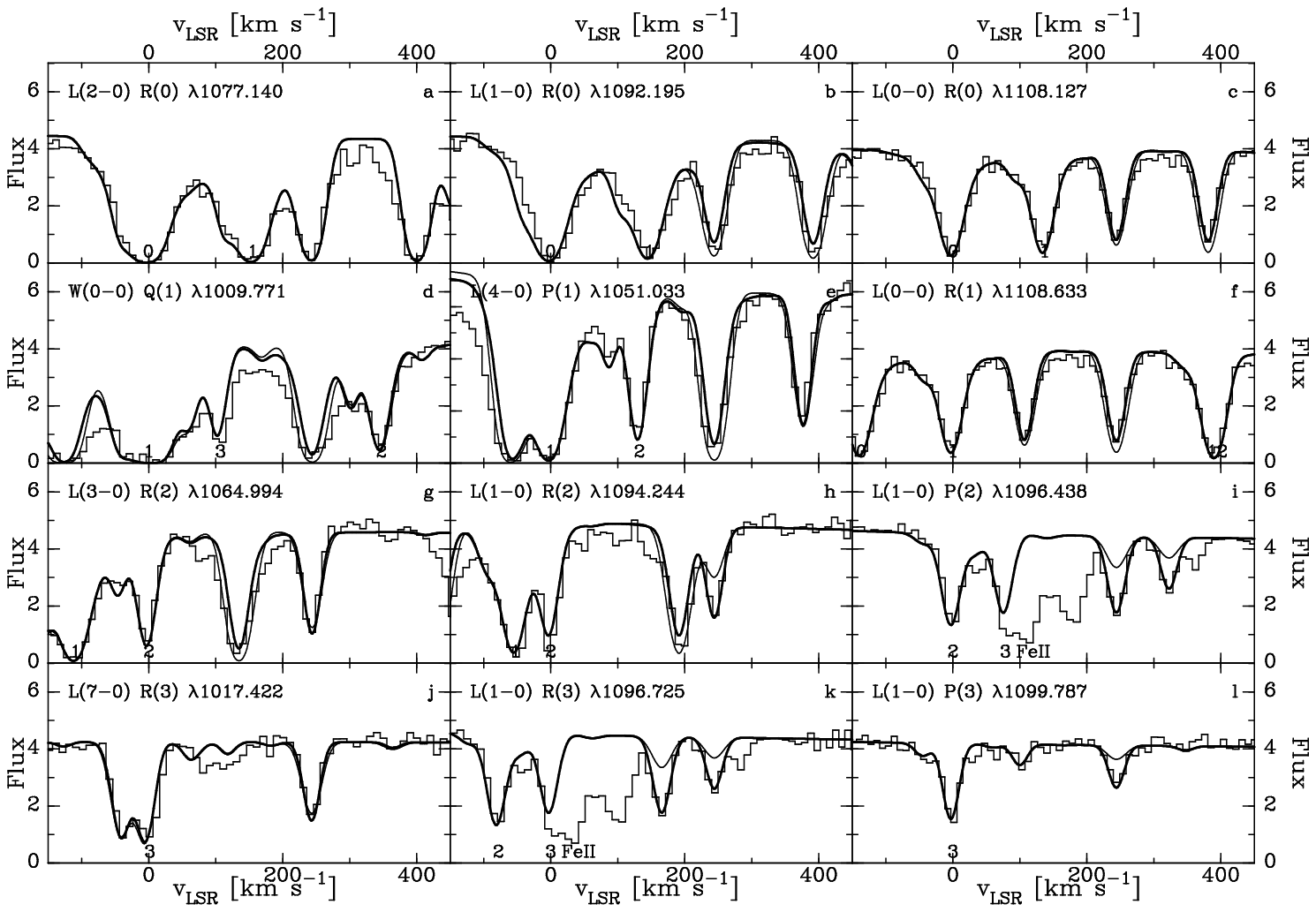}\figurenum{3}\caption{\captionC}\end{figure}
\newpage\par\begin{figure}\plotfiddle{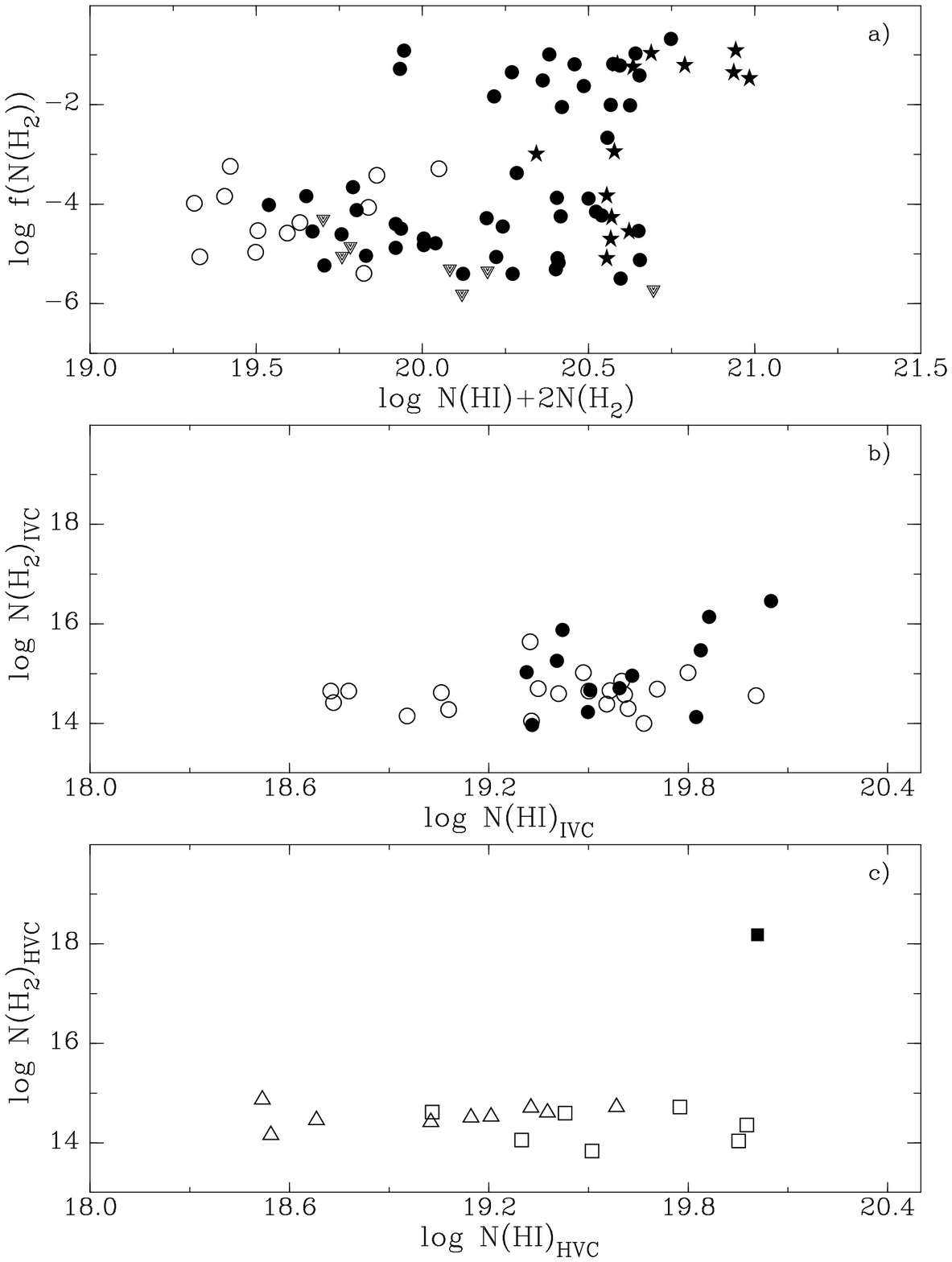}{0in}{0}{345}{490}{0}{00}\figurenum{4}\caption{\captionD}\end{figure}
%newpage\par\begin{figure}\plotone{f4.eps}\figurenum{4}\caption{\captionD}\end{figure}
\newpage\par\begin{figure}\plotone{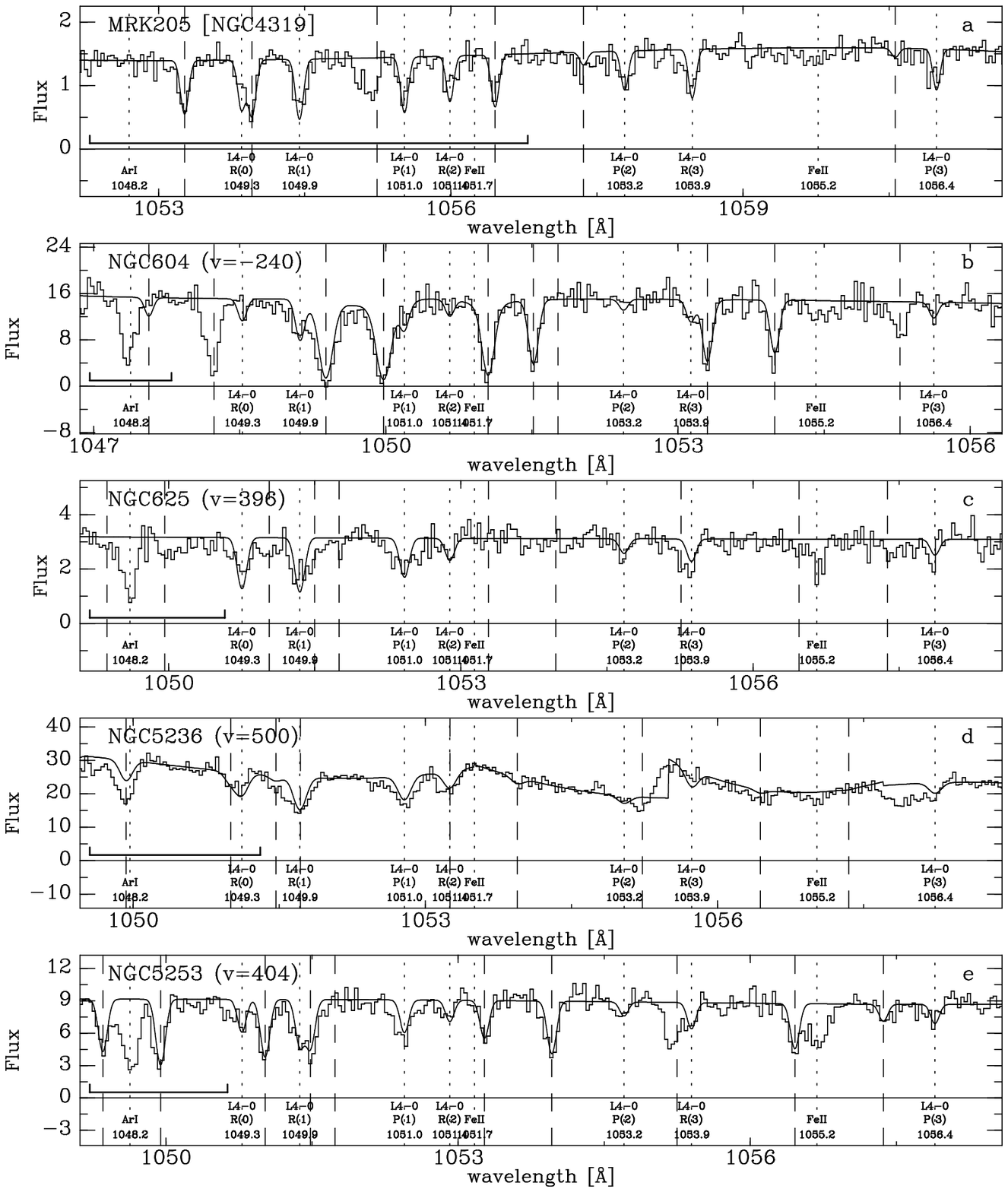}\figurenum{5}\caption{\captionE}\end{figure}
\newpage\par\begin{figure}\plotone{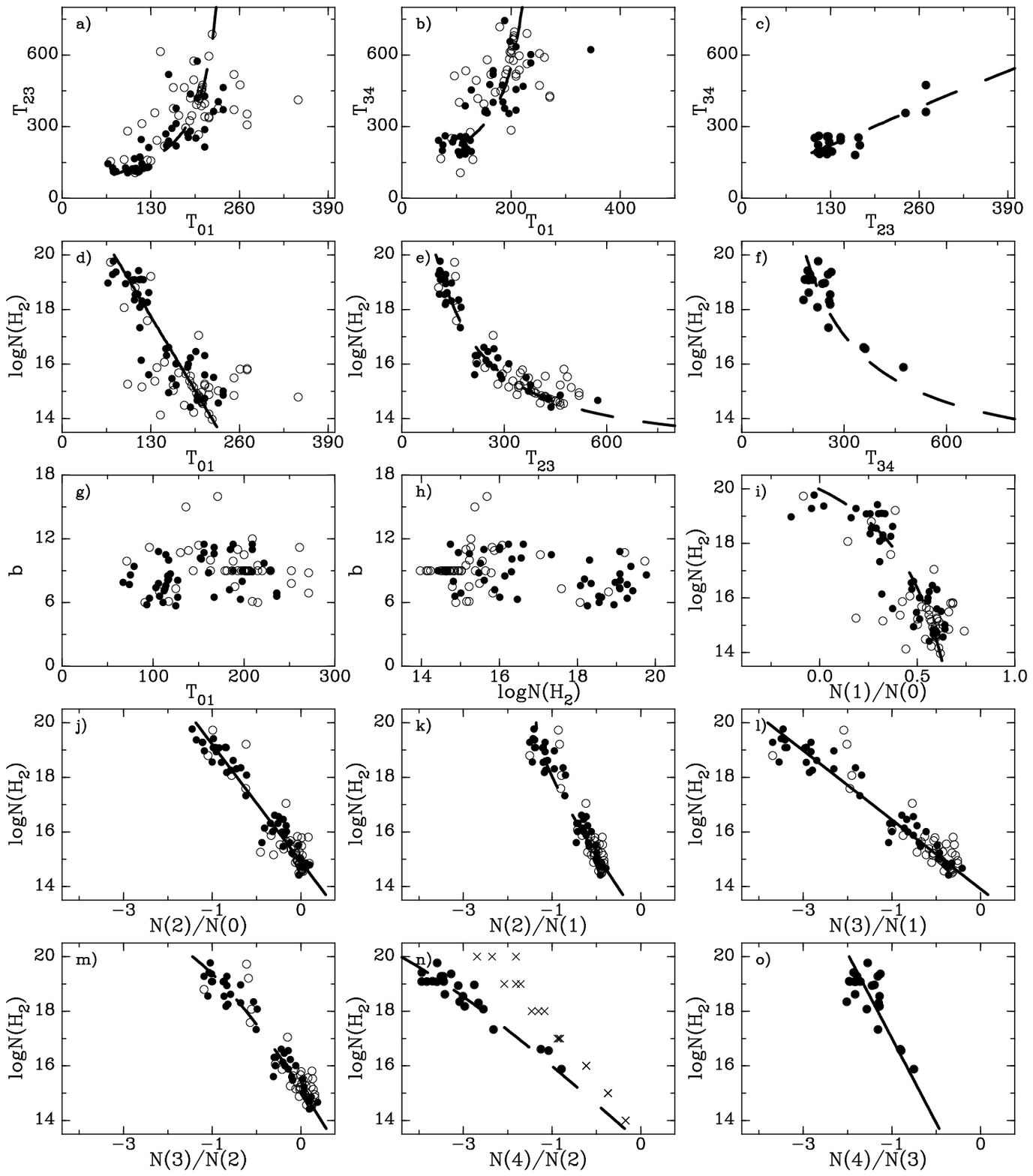}\figurenum{6}\caption{\captionF}\end{figure}
%\newpage\par\begin{figure}\plotone{f7.eps}\figurenum{7}\caption{\captionG}\end{figure}
%\newpage\par\begin{figure}\plotone{f8.eps}\figurenum{8}\caption{\captionH}\end{figure}

%\documentclass[11pt,preprint]{aastex} \begin{document} \pagestyle{empty}
\hoffset-2cm
\def\kms{km\,s$^{-1}$}
\def\diffnomA{1}
\def\diffnomB{2}
\def\MDRS{3}
\def\sl{\hbox to 0pt{$^4$\hss}}
\begin{deluxetable}{lrrlllcrrrrrr}
\tabletypesize{\footnotesize}
\tabcolsep=3pt \tablenum{1} \tablewidth{0pt}

\tablecolumns{13}
\tablecaption{Target Exposure List}
\tablehead{\colhead{object} &   \colhead{lon}  &   \colhead{lat}  & \colhead{type} & \colhead{dataset} & \colhead{Cal}  & \colhead{T$_{\rm exp}$} & \multicolumn{6}{c}{shifts}                                            \\
              &\colhead{[$\circ$]}&\colhead{[$\circ$]}&        &             & \colhead{FUSE} & \colhead{[ks]}          & \colhead{LiF1A} & \colhead{LiF1B} & \colhead{LiF2A} & \colhead{LiF2B} & \colhead{SiC2A} & \colhead{SiC2B} \\
   \colhead{(1)}    &   \colhead{(2)}  &   \colhead{(3)}  & \colhead{(4)}  & \colhead{(5)}     & \colhead{(6)}  & \colhead{(7)}           & \colhead{(8)}   & \colhead{(9)}   & \colhead{(10)}  & \colhead{(11)}  & \colhead{(12)}  & \colhead{(13)}   }
\startdata
1H0707$-$495           & 260.17 & $-$17.67 & Sey1       & B1050101 & 2.4 & 20.3 & $-$13 & $-$13 & $-$26 & $-$26 & $-$13 & $-$13\\
                     &        &        &            & B1050102 & 2.4 & 24.2 & $-$19 & $-$19 & $-$19 & $-$19 & $-$13 & $-$13\\
                     &        &        &            & B1050103 & 2.4 & 22.5 & $-$19 & $-$19 & $-$19 & $-$19 & $-$13 & $-$13\\
1H0717+714           & 143.98 &  28.02 & BLLac      & Z9071301 & 2.4 & 47.6 & 14 & 14 & 3 & 14 & 3 & $-$8\\
3C249.1              & 130.39 &  38.55 & QSO        & P1071601 & 2.1 & 11.4 & 16 & 16 & 7 & 7 & 50 & 50\\
                     &        &        &            & P1071602 & 2.1 & 9.5 & 25 & $-$ & $-$ & $-$ & 50 & 50\\
                     &        &        &            & S6010901 & 2.1 & 30.7(25.7$^{\diffnomA}$) & 9 & $-$ & $-$5 & $-$5 & $-$5 & $-$5\\
                     &        &        &            & P1071603 & 2.4 & 80.9 & 14 & 14 & 14 & 14 & 4 & 4\\
\enddata
\tablecomments{
See Sects.~2.1 and 2.2 for a detailed description of the columns;
\diffnomA: Useable exposure time in LiF1A segment $>$10\% less than nominal time;
\diffnomB: Useable exposure time in LiF2A segment $>$10\% less than nominal time;
\MDRS: Observed in the MDRS aperture, rather than in the LWRS aperture;
4: Observations for which some segments show a clear slope in the centroid vs wavelength plot; specifically (in units of km\,s$^{-1}$/\AA):
ESO141-G55--I9040104 (1999 10 16): 1alif=$-$0.1206, 1blif=$-$0.2395, 2alif=$-$0.2336, 2blif=$-$0.1094;
H1821+643--P1016405 (1999 10 13):  1alif=$-$0.1108, 2blif=$-$0.1643;
Mrk335--P1010204 (2000 11 21):     1alif=0.1179, 2blif=$-$0.1113;
Mrk509--X0170101 (1999 11 02):     1alif=$-$0.1539, 1blif=$-$0.1590, 2alif=$-$0.2923, 2blif=$-$0.1565;
Mrk876--P1073101 (1999 10 16):     1alif=$-$0.1383, 2blif=$-$0.1597;
MS0700.7+6338--P2072701 (2000 01 10): 1alif=0.1721;
NGC604--B0180201 (2001 09 03):     1alif=$-$0.1876;
PG0804+761--P1011901 (1999 10 05): 1alif=$-$0.1691, 2blif=$-$0.1149;
VIIZw118--P1011604 (1999 10 06):   1alif=$-$0.1149, 2blif=$-$0.1377.
}
\end{deluxetable}
%\end{document}

%\documentclass[11pt,preprint]{aastex} \begin{document} \pagestyle{empty}
\hoffset-2cm
\def\kms{km\,s$^{-1}$}
\def\notefu{1}
\def\FO{\null}\def\FF{\hbox to 0pt{$^a$}}
\def\notemrkfn{2}
\def\notengctsno{3}
\def\notengcfffn{4}
\def\notengcfsso{5}
\begin{deluxetable}{lrrrrrrrcr}
\tabletypesize{\footnotesize}
\tabcolsep=3pt \tablenum{2} \tablewidth{0pt}

\tablecolumns{10}
\tablecaption{S/N ratios}
\tablehead{\colhead{object} & \colhead{Flux}       & \colhead{S/N}  & \colhead{S/N}  & \colhead{S/N} & \colhead{S/N}  & \colhead{$v$(H\,I)} & \colhead{$v$(metal)} & \colhead{$v$(H$_2$)} & \colhead{instr} \\
              & \colhead{[f.u.]$^1$} & \colhead{1031} & \colhead{1063} & \colhead{977} & \colhead{1122} & \colhead{[\kms]}    & \colhead{[\kms]}     & \colhead{[\kms]}     &           \\
   \colhead{(1)}   &  \colhead{(2)}       &  \colhead{(3)}  &  \colhead{(4)} & \colhead{(5)} & \colhead{(6)}  & \colhead{(7)}       & \colhead{(8)}        & \colhead{(9)}        & \colhead{(10)}   }
\startdata
1H0707$-$495           &  1.76\FO    &~11.66 & 12.02 &  4.04 &  8.27 & $-$1,61 & 2,61 & $-$1 & 21cm$-$H2 \\
1H0717+714           &  2.70\FO    &~11.69 & 12.69 &  4.57 & 12.25 & $-$51,$-$5 & $-$51,$-$5 & $-$51,$-$5 & 21cm \\
3C249.1              &  1.10\FO    &~13.75 & 12.12 &  3.36 & 12.68 & $-$50,9 & $-$47,$-$2 & 3 & E140M \\
3C273.0              & 26.86\FO    &~31.16 & 25.38 & 19.24 & 39.03 & $-$6,25 & $-$14,22 & $-$9,28 & E140M \\
ESO141$-$G55           &  5.17\FO    &~15.96 & 22.96 &  7.78 & 19.96 & $-$45,$-$1 & $-$1 & $-$4 & 21cm \\
ESO572$-$G34           &  6.39\FF    &~11.21 & 13.83 &  4.51 & 14.58 & $-$4,0 & $-$4 & $-$4 & 21cm \\
H1821+643            &  3.04\FO    &~27.89 & 23.97 &  7.47 & 26.09 & $-$128,$-$87,$-$3 & $-$129,$-$82,$-$13 & $-$5 & E140M \\
HE0226$-$4110          &  2.73\FO    &~24.59 & 21.23 & 10.97 & 18.41 & $-$7,0 & $-$4 & $-$7 & E140M \\
\enddata
\tablecomments{
See Sect~2.2 for a detailed description of the columns;
\notefu: flux in units of 10$^{-14}$\,erg\,cm$^{-2}$\,s$^{-1}$\,\AA$^{-1}$ --
an (a) indicates the flux value is for 1063\,\AA, instead of 1031\,\AA;
\notemrkfn:   average between $-$20 and $-$1~\kms;
\notengctsno: uncertain velocity -- broad ISM lines without easily recognizable components;
\notengcfffn: average between $-$11 and $-$45~\kms;
\notengcfsso: average between $-$26 and $-$8~\kms;
}
\end{deluxetable}
%\end{document}

%\documentclass[11pt,preprint]{aastex} \begin{document} \pagestyle{empty}
\hoffset-2cm
\def\kms{km\,s$^{-1}$}

\begin{deluxetable}{lrrrrrrrccccc}
\tabletypesize{\footnotesize}
\tabcolsep=3pt \tablenum{3} \tablewidth{0pt}

\tablecolumns{13}
\tablecaption{H$_2$ measurements}
\tablehead{\colhead{object} & \colhead{$v$(H$_2$)} & \colhead{$N$(J=0)} & \colhead{$N$(J=1)} & \colhead{$N$(J=2)} & \colhead{$N$(J=3)} & \colhead{$N$(J=4)} & \colhead{$N^a$(H$_2$)}& \colhead{$W^b$} &\colhead{$W_f^c$} &\colhead{$T_{01}^d$}&\colhead{$T_{23}^d$}&\colhead{$T_{34}^d$}\\
              & \colhead{[\kms]}     &              &              &              &              &              &                 & \colhead{[\kms]}&\colhead{[\kms]}  &\colhead{[K]}       &\colhead{[K]}       &\colhead{[K]}       \\
   \colhead{(1)}    & \colhead{(2)}        & \colhead{(3)}      & \colhead{(4)}      & \colhead{(5)}      & \colhead{(6)}      & \colhead{(7)}      & \colhead{(8)}         & \colhead{(9)}   &\colhead{(10)}    &\colhead{(11)}      &\colhead{(12)}      &\colhead{(13)}       }
\startdata

1H0707$-$495           &$-1$ &  $14.73_{-0.26}^{+0.41}$  &  $15.29_{-0.38}^{+0.71}$  &  $14.68_{-0.26}^{+0.34}$  &  $14.78_{-0.25}^{+0.36}$  &  $<$14.56 &   $15.56_{-0.18}^{+0.51}$ &   $9.9_{-1.9}^{+5.0}$    & lin   &  188 &   420 &   (462)\\

                     &$61$ &  $<$14.35  &  $<$14.48  &  $<$14.50  &  $<$14.50  &  $<$14.44 &   $<$14.59 &   (9.0)    & --   &      &       &      \\

1H0717+714           &$-51$ &  $14.12_{-0.18}^{+0.16}$  &  $14.71_{-0.23}^{+0.22}$  &  $<$14.56  &  $<$14.52  &  $<$14.52 &   $14.96_{-0.17}^{+0.18}$ &   $9.0_{-0.0}^{+0.0}$    & lin   &  204 &   (387) &   (577)\\

                     &$-5$ &  $16.28_{-0.27}^{+0.74}$  &  $16.87_{-0.40}^{+0.76}$  &  $16.06_{-0.30}^{+0.78}$  &  $15.87_{-0.28}^{+0.77}$  &  $14.48_{-0.14}^{+0.19}$ &   $17.04_{-0.22}^{+0.63}$ &   $10.5_{-2.5}^{+2.5}$    & --   &  200 &   267 &   285\\

3C249.1              &$-50$ &  $<$15.13  &  $<$15.72  &  $<$14.30  &  $<$14.27  &  $<$14.20 &   $<$15.01 &   (9.0)    & --   &      &       &      \\

                     &$3$ &  $18.58_{-0.05}^{+0.07}$  &  $18.91_{-0.05}^{+0.05}$  &  $17.48_{-0.14}^{+0.16}$  &  $16.34_{-0.17}^{+0.12}$  &  $14.57_{-0.21}^{+0.20}$ &   $19.09_{-0.03}^{+0.03}$ &   $8.7_{-0.5}^{+0.5}$    & --   &  119 &   125 &   207\\

3C273.0              &$-9$ &  $13.94_{-0.11}^{+0.10}$  &  $14.58_{-0.11}^{+0.13}$  &  $14.01_{-0.10}^{+0.10}$  &  $14.16_{-0.09}^{+0.10}$  &  $<$13.79 &   $14.86_{-0.06}^{+0.07}$ &   $6.6_{-0.8}^{+1.1}$    & 6.0   &  236 &   464 &   (602)\\

                     &$28$ &  $15.14_{-0.12}^{+0.12}$  &  $15.69_{-0.11}^{+0.01}$  &  $14.90_{-0.07}^{+0.07}$  &  $14.71_{-0.06}^{+0.07}$  &  $13.73_{-0.13}^{+0.10}$ &   $15.88_{-0.07}^{+0.02}$ &   $7.2_{-0.2}^{+0.2}$    & 7.4   &  184 &   270 &   474\\
\enddata
\tablecomments{
a: FWHM values in Col.~9 that are in parentheses imply an FWHM of 9~\kms\ was used to convert the equivalent width limits into column density limits.
b: $W_f$ is the FWHM fitted by combining all H$_2$ lines without damping wings, rather than by assuming that $T_{12}=T_{23}$.
c: Temperature values in parentheses are not measured from observed column density ratios, but are expected values based on the correlations shown in Fig.~6.
d: The listed errors for NGC\,5236 and NGC\,5253 do not include the systematic errors associated with the uncertain continuum placements.
e: Sightlines where lines of $J$=5 are detected:
ESO\,141$-$G55 -- log\,$N$(5)=14.32$_{-0.22}^{+0.17}$;
HS\,0624+6907  -- log\,$N$(5)=14.67$_{-0.25}^{+0.35}$;
Mrk\,116       -- log\,$N$(5)=14.47$_{-0.21}^{+0.22}$;
Mrk\,876       -- log\,$N$(5)=13.94$_{-0.12}^{+0.08}$;
Mrk\,1095      -- log\,$N$(5)=14.53$_{-0.20}^{+0.18}$;
NGC\,3783      -- log\,$N$(5)=14.24$_{-0.08}^{+0.10}$;
NGC\,4151      -- log\,$N$(5)=13.45$_{-0.25}^{+0.16}$;
NGC\,5253      -- log\,$N$(5)=14.23$_{-0.14}^{+0.13}$;
NGC\,7714      -- log\,$N$(5)=14.17$_{-0.32}^{+0.25}$;
PG\,1211+143   -- log\,$N$(5)=13.98$_{-0.11}^{+0.12}$;
PHL\,1811      -- log\,$N$(5)=14.36$_{-0.15}^{+0.12}$;
PKS\,0405$-$12 -- log\,$N$(5)=14.05$_{-0.19}^{+0.15}$
}
\end{deluxetable}
%\end{document}

%\documentclass[11pt,preprint]{aastex} \begin{document} \small \pagestyle{empty}
%TABLE4
\begin{deluxetable}{llclclcl}
\tabcolsep=4pt \tablenum{4} \tablewidth{0pt} \tablecolumns{8}
\tabletypesize{\footnotesize}
\tablecaption{H$_2$ outside the Milky Way Disk}
\tablehead{\colhead{object}&\colhead{cloud name}&\colhead{}            &\multicolumn{2}{c}{Published value}      &\multicolumn{2}{c}{This paper}          &\colhead{Note} \\
           \colhead{}      &\colhead{}          &\colhead{$v$(IVC)}    &\colhead{log\,$N$(H$_2$)}&\colhead{FWHM} &\colhead{log\,$N$(H$_2$)}&\colhead{FWHM}&\colhead{}     \\
           \colhead{}      &\colhead{}          &\colhead{km\,s$^{-1}$}&\colhead{}        &\colhead{km\,s$^{-1}$}&\colhead{}       &\colhead{km\,s$^{-1}$}&\colhead{}      }
\startdata
\sidehead{Intermediate-velocity H$_2$ (compared to Richter et al.\ 2001).}\\
1H\,0717+714 & LLIV Arch   &  $-$51   &                    &      & 14.96$\pm$0.18  & 9.0      & new sightline         \\ %OK         newlos
3C\,273.0    &             &    +25   & 15.71              & 10.0 & 15.82$\pm$0.05  & 7.5      & similar               \\ %OK sim
Mrk\,59      & IV Arch     &  $-$44   & 14.72$\pm$0.28     &  6.8 & 15.03$\pm$0.14  & linear   & similar               \\ %OK sim
Mrk\,153     &             &          &                    &      & 15.83$\pm$0.25  & 10.5     & new sightline         \\ %OK         newlos
Mrk\,279     & LLIV Arch   &  $-$41   & $<$14.85           &      & 14.42$\pm$0.09  & linear   & additional data       \\ %OK     addl
Mrk\,421     & IV 26       &  $-$60   & $<$14.93           &      & 14.12$\pm$0.18  & linear   & additional data       \\ %OK     addl
Mrk\,509     & complex gp  &    +60   & 14.92$\pm$0.46     &  9.2 & 14.67$\pm$0.06  & 9.0      & similar               \\
\enddata
\end{deluxetable}
%\end{document}
%TABLE4

\end{document}